\documentclass[useAMS,usenatbib,usegraphicx]{mn2e}

\usepackage{amsmath}
\usepackage{bm}
\usepackage{graphicx}

\newcommand{\beq}{\begin{equation}}
\newcommand{\eeq}{\end{equation}}
\newcommand{\bea}{\begin{eqnarray}}
\newcommand{\eea}{\end{eqnarray}}
\newcommand{\req}[1]{Eq.~(\ref{#1})}
\newcommand{\low}{^{\phantom{}}}
\newcommand{\alphaf}{\alpha_\mathrm{f}} % fine-structure constant
\newcommand{\dd}{\mathrm{d}} % differential
\newcommand{\gcc}{\mbox{g cm$^{-3}$}}
\newcommand{\mel}{m_e} % electron mass
\newcommand{\mpr}{m_p} % proton mass
\newcommand{\am}{a_\mathrm{m}}
\newcommand{\beti}{{\beta_\mathrm{i}}}
\newcommand{\betp}{{\beta_p}}
\newcommand{\Bq}{B_{\rm Q}}
\newcommand{\dir}{\mathbf{n}}

\newcommand{\Gamrp}{\Gamma_{\mathrm{r},p}}
\newcommand{\gtrsim}{\ga}
\newcommand{\hi}{\chi^{\phantom{|_|}}}
\newcommand{\ks}{k_\mathrm{s}}
\newcommand{\lesssim}{\la}
\newcommand{\mion}{m_\mathrm{i}}
\newcommand{\nion}{n_\mathrm{i}}
\newcommand{\npr}{n_p}
\newcommand{\omc}{\omega_{\mathrm{c}}} % generic cyclotron
\newcommand{\omce}{\omega_{\mathrm{c}e}} % electron cyclotron
\newcommand{\omcp}{\omega_{\mathrm{c}p}} % proton cyclotron
\newcommand{\omci}{\omega_{\mathrm{ci}}} % ion cyclotron
\newcommand{\pe}{{(pe)}}
\newcommand{\pp}{{(pp)}}
\newcommand{\ppx}{{(pp,x)}}
\newcommand{\us}{u_\mathrm{s}}

\begin{document}

\title[Statistical equilibrium in magnetized plasmas]{Statistical 
equilibrium and ion cyclotron absorption/emission
 in strongly magnetized plasmas}

\author[Potekhin \& Lai]{
Alexander Y. Potekhin$^{1,2}$
and 
Dong Lai$^{2}$
\\
$^{1}$Ioffe Physico-Technical Institute,
  Politekhnicheskaya 26, 194021 St.~Petersburg, Russia;
{palex@astro.ioffe.ru}
\\
$^{2}$Center for Radiophysics and Space Research, 
 Department of Astronomy,
 Cornell University,
 Ithaca, NY 14853, USA}

\maketitle

\begin{abstract}
We calculate the transition rates between proton Landau levels 
due to non-radiative and radiative Coulomb
collisions in an electron-proton plasma with strong magnetic field $B$.
Both electron-proton collisions and proton-proton collisions
are considered.
The roles of the first-order cyclotron absorption and second-order free-free
absorption and scattering in determining the line strength and shape
as well as the continuum are analysed in detail. We 
solve the statistical balance equation for the populations of proton Landau 
levels. For temperatures $\sim10^6$\,--\,$10^7$~K,
the deviations of the proton populations from LTE are appreciable
at density $\rho\lesssim0.1 B_{14}^{3.5}$ \gcc,
where $B_{14}=B/(10^{14}~{\rm G})$.
We present general formulae for the plasma emissivity and absorption
coefficents under a wide range of physical conditions.
Our results are useful for studying the possibility and the conditions 
of proton/ion cyclotron line formation in the near vicinity of
highly magnetized neutron stars.
\end{abstract}

% ************************************************************
% TEXT BODY
% ************************************************************

\section{Introduction}
\label{sect:intro}

Cyclotron lines are a powerful diagnostic tool for magnetized neutron 
stars. The detection of electron cyclotron features at $10$\,--\,80~keV
in the spectra of a number of binary X-ray pulsars 
(e.g., \citealt{Trumper-ea78}; see \citealt{Heindl-ea04},
\citealt{Terada-ea06}
for recent 
observations) provided direct confirmation that these are 
neutron stars endowed with strong magnetic fields $B\sim 10^{12}$\,--\,$10^{13}$~G.
Numerous theoretical works have been devoted to the physics and modelling
of electron cyclotron line formation and transfer in accreting neutron 
stars
\citep*[e.g.,][]{WassermanSalpeter,MN85,BKA88,LWW90,WWL93,ArayaHarding,ArayaG-Harding}.

There has been growing evidence in recent years for the existence of neutron 
stars possessing superstrong magnetic fields, $B\gtrsim 10^{14}$~G.
In particular, soft gamma-ray repeaters (SGRs) and anomalous X-ray
pulsars (AXPs) are believed to be magnetars, whose radiation
is powered by the decay of superstrong magnetic fields 
\citep[see][]{ThompsonDuncan05,ThompsonDuncan06,WoodsThompson}. 
Several radio pulsars with inferred surface magnetic fields
approaching $10^{14}$~G have also been discovered (e.g., 
\citealt{McLaughlin-ea03}). 
In such superstrong magnetic field regime, the 
electron cyclotron energy, 
\beq
\hbar\omce=
\hbar {eB\over m_ec}=1.16~B_{14}~{\rm MeV},
     \label{e-cycl}
\eeq
lies outside the X-ray band, 
but the
     ion
 cyclotron energy,
\beq
     \hbar\omci=\hbar{ZeB\over \mion c}=0.635\,(Z/A) B_{14}~{\rm keV},
     \label{i-cycl}
\eeq
lies in the detectable range for X-ray telescopes such as 
{\it Chandra} and {\it XMM-Newton} when $B_{14}\gtrsim 0.4$. 
In Eqs.~(\ref{e-cycl}) and (\ref{i-cycl}), 
   $B_{14}=B/(10^{14}~{\rm G})$,
   $\mion$ is an ion mass,
    and $Z$ and $A$ are nuclear charge and mass numbers.
In the last few years, absorption features at $E\sim 0.2$\,--\,1 keV
have been detected in the spectra of several thermally emitting isolated
neutron stars
\citep[e.g.,][]{Haberl-ea04,Kerkwijk-ea04,Kerkwijk06}.
While no definitive identifications of the lines have been made,
it is likely that some of these lines involve proton cyclotron resonances
at $B\lesssim 10^{14}$~G. Somewhat surprisingly, the observed quiescent 
emission of AXPs and SGRs
does not show any spectral feature, in particular the proton cyclotron line 
around 1~keV
 \citep[e.g.,][]{Juett-ea02,Patel-ea03,Kulkarni-ea03,Tiengo-ea}.
This absence of lines may be naturally explained by the effect of vacuum polarization, 
which tends to reduce the line width 
significantly in the atmosphere (thermal) emission for $B\gtrsim 10^{14}$~G
\citep{HoLai03,HoLai04,LaiHo03,vanAdelLai06}.

There has been some evidence for
     ion
 cyclotron lines during 
several AXP/SGR outbursts, e.g., the 6.4~keV emission feature in SGR 1900+14 
\citep{StrohmayerIbrahim}, the 5~keV absorption feature in SGR 1806$-$20 
\citep*{ISP03}, and the 14~keV emission feature (and possibly also 
$\sim 7,~30$~keV features) in the bursts of AXP 1E~1048$-$5937 
\citep{GKW02}. There was also a possible detection of 
a 8.1~keV absorption feature in AXP 1RXS J1708$-$4009
 (\citealt{Rea-ea03};
but see \citealt{Rea-ea05}). It is possible that these 
absorption/emission features are produced by 
     ion
 cyclotron 
resonances in the corona or
lower magnetosphere of magnetars
     (although one cannot exclude 
     the alternative possibility that they are produced by electron cyclotron
     resonances in upper magnetospheres).

     To be specific, in the following we focus on the electron-proton plasma
     ($Z=1$, $A=1.008$, spin $=\frac12$)
     and the proton cyclotron resonance with energy
     $\hbar\omcp=0.630\,B_{14}$ keV.
     Generalization to other ions is outlined in Sect.~\ref{sect:Z}.

While the physics of electron cyclotron line transfer has been
extensively studied in the context of accreting X-ray pulsars
(see above), several basic issues regarding proton cyclotron line 
have not been properly considered.
Since the radiative electron cyclotron decay rate is
many orders of magnitude larger than the collisional deexcitation
rate, electron cyclotron resonance always takes the form of scattering
\citep[e.g.][]{Mesz}.
For protons, the radiative cyclotron decay rate
is much smaller, so the situation is not at all clear.  Depending on the plasma
density, temperature and magnetic field strength, true proton
cyclotron absorption and emission are possible.
Previous calculations of proton cyclotron lines from magnetized neutron star 
atmospheres 
\citep[e.g.,][]{Zane-ea01,HoLai01,HoLai03,PLCH04}
     assumed local thermal equilibrium (LTE) population of proton
     Landau levels. We will see that this assumption is not always
     valid in the case of a magnetar.

In this paper we study systematically the rates for collision-induced
proton cyclotron transitions in a magnetized plasma. Combining these
rates with radiative transition rates, we then study 
the statistical equilibrium of protons in different Landau 
levels, and use the non-LTE level population to calculate the 
radiative opacities and emissivities for different photon modes.
Our results serve as an crucial ingredient for determining the possibility and 
the physical conditions of proton/ion cyclotron line formation in various
plasma environments of highly magnetized neutron stars.

In statistical equilibrium, populations of excited 
Landau levels of the ions 
are determined by 
rates of spontaneous radiative decay
and by rates of transitions caused by
radiative and non-radiative Coulomb collisions
(note that `true' cyclotron absorption can be separated 
from scattering by considering second-order Feynman diagrams 
which include Coulomb interaction; cf.\ \citealt{DV78}).
Order-of-magnitude estimates of the rates of these processes
and their consequences for the level populations are given 
in Sect.~\ref{sect:ordermag}. 
In subsequent sections we consider these processes in more detail.
In Sect.~\ref{sect:Coul} we write formulae 
for rates of such transitions in a proton-electron plasma.
Radiative transition rates and cross sections 
are considered in Sect.~\ref{sect:Rad}.
In Sect.~\ref{sect:stateq} we analyse population of the Landau states
of the ions.
Opacity and emissivity for the ion cyclotron resonance
are calculated in Sect.~\ref{sect:opem}.

%********************************
\subsection{Landau Level Basics and Notations}
\label{sect;notations}

Motion of the ions and electrons in the plane ($xy$) perpendicular 
to the magnetic field $\bm{B}$ (assumed to  be directed along $z$)
is quantized in Landau levels with excitation energies 
$E_{N,\perp} = mc^2\,[\sqrt{1+2bN}-1]$ ($N=0,1,\ldots$), 
where $b=\hbar\omc/mc^2$ is the relativistic magnetic parameter,
$\omc = |Ze| B / m c$ is the cyclotron frequency, and
$Ze$ and $m$ are the charge and mass of the particle
($e$ is the elementary charge).
For non-interacting particles, 
every Landau level is degenerate with respect to a position of the 
guiding centre in the $xy$ plane
 (e.g., \citealt{LaLi-QM}). The number of degenerate states
(for fixed $N$ and longitudinal velocity $v_z$)
is $L_x L_y |Z| /(2\pi \am^2)$, where 
\beq
\am=\left({\hbar c\over eB}\right)^{1/2}
=2.5656\times 10^{-11}B_{14}^{-1/2}~{\rm cm}
\label{eq:am}\eeq
is the magnetic length, 
 and $L_x$ and $L_y$ are normalization lengths.
In addition, excited Landau levels of the electrons
usually can be treated as double spin-degenerate.
In contrast, the Landau levels of the ions are not degenerate, but split
with respect to the spin, 
because of the anomalous magnetic moments of nuclei.

For the electrons,  the relativistic magnetic parameter is
$b_e =\hbar\omce/(m_ec^2)= B/\Bq$, 
where 
$\Bq=m_e^2c^3/(e\hbar)=4.414\times10^{13}$~G.
For density $\rho\ll 7\times 10^6\,
     (A/Z)
B_{14}^{3/2}$~g~cm$^{-3}$
and temperature 
$T\ll \hbar\omce/k_B=1.34\times 10^{10}~B_{14}$~K, 
virtually all electrons reside in the ground Landau level.
For $T\gtrsim 2.7\times 10^{-4}~B_{14}^{-2}(\rho_0 Z/A)^2$~K (where $\rho_0$
is the density in units of 1~g~cm$^{-3}$), the electrons are non-degenerate.
We shall be concerned with this regime in this paper.

For the ions, the cyclotron energy is
$\hbar\omci= 0.635\, (Z/A)\,B_{14}$~keV,
and their relativistic magnetic parameter is 
$b_\mathrm{i} = \hbar\omci/(\mion c^2)=
b_e Z (\mel/\mion)^2
  = 0.68\times10^{-6}\,(Z/A^2)\, B_{14} .
$
We consider the situation where the ions are non-relativistic:
$Nb_\mathrm{i} \ll 1$, $E_{N,\perp}=N\hbar\omci$.
Obviously it is always the case if the ion Landau number $N$
is not huge.

In the following we introduce a number of notations for
various kinds of transition rates (i.e., number of transitions
per unit time per 
occupied initial state) 
between proton Landau 
levels $N$ and $N'$ and corresponding cross sections. Here we list these
and other related 
notations and the basic relevant equations for easy reference:

\smallskip
\halign{#\hfil&~\vtop{\parindent=0pt\hsize=2.2in\strut#\strut}\cr
$\omce$, $\omci$, $\omcp$
  & cyclotron frequencies for electron, ion and proton: Eqs.~(\ref{e-cycl}),
    (\ref{i-cycl})\cr
$\am$
  & magnetic length: \req{eq:am}\cr
$n_e$, $\nion$, $\npr$
  & number densities of electrons, ions and protons in the plasma\cr
$n_N$
  & number density of protons in the $N$-th Landau level ($N=0,1,2,\cdots$)\cr
$W_{NN'}^\mathrm{fix}(v_z)$ 
  & transition rate per particle
    on the level $N$ 
    with initial longitudinal (i.e., along $\bm{B}$) velocity $v_z$,
    in a volume $V$
    for scattering on a fixed Coulomb centre: \req{WNN'fix}\cr
$\sigma_{NN'}^\mathrm{fix}(v_z)$ 
  & corresponding cross section:
    Eqs.~(\ref{WNN'fix}), (\ref{WNN'fixZ})\cr
$W^\pe_{NN'} (v_z)$
  & transition rate for protons scattered on those electrons which rest
    on the ground Landau level, assuming the relative 
    initial longitudinal velocity $v_z$: \req{Wpe}\cr
$\sigma^\pe_{NN'}(v_z)$ 
  & partial cross section for such scattering, normalized to $v_z$:
    Eqs.~(\ref{Gamma10}), (\ref{Wpe})\cr
$W_{{N_1 N_2};\,{N_1' N_2'}}(v_z)$
  & transition rate for charged particle 1 scattered on particle 2
    in volume $V$ at relative longitudinal velocity $v_z$,
    under the condition that initial and final Landau numbers
    of particle $i$ are $N_i$ and $N'_i$, respectively:
    \req{WN1N2}\cr
$W_{N_1,N_2;\,N_1',N_2'}^\pm(v_z)$
  & analogous to $W_{{N_1 N_2};\,{N_1' N_2'}}(v_z)$,
    but for collision of identical particles with even ($+$)
    or odd ($-$) total spin: \req{W-pm}\cr
$\sigma_{NN_2;\,N' N_2'}^{\pp\pm}$ 
  & corresponding cross section normalized to $v_z$:
    Eqs.~(\ref{Gpp}), (\ref{gam-pp}), (\ref{WN1N2})\cr
$\sigma_{NN'}^{(N_2)}$
  & partial cross section of a proton with initial Landau number $N_2$
    (regardless of its final Landau number)
    with respect to scattering of a test proton from Landau number $N$
    to $N'$: Eq.~(\ref{GppL})\cr
$\Gamma^{C\pe}_{NN'}$
  & velocity-averaged partial Coulomb transition rate
    for proton scattering on the electrons on the ground Landau level:
    Eqs.~(\ref{Gamma10}), (\ref{GamAtoC}), (\ref{Gam-C})\cr
$\Gamma^{C\pp}_{NN'}$
  & analogous to $\Gamma^{C\pe}_{NN'}$, but because of collisions 
    with other protons: Eqs.~(\ref{Gpp}), (\ref{GppL})\cr
$\Gamma^A_{NN'}$
  & spontaneous decay rate: Eqs.~(\ref{GammA10}), (\ref{Gam-A}), 
    (\ref{GammA})\cr
$\Gamma^B_{NN'}$
  & total transition rate due to photoabsorption: 
    Eqs.~(\ref{Gam-B}), (\ref{Gamma01})\cr
$\hat\Gamma^B_{NN'}$
  & total transition rate due to stimulated emission: \req{Gamma01}\cr
$\sigma_{NN'}(j,\omega,\dir)$
  & cross section of a proton with respect 
    to photoabsorption (normalized to speed of light)
    for polarization $j$ ($j=1,2$), photon frequency $\omega$ and 
    direction $\dir$:
    Eqs.~(\ref{BbarJ}), (\ref{sumalpha}), (\ref{GamA_NN}), (\ref{muNN'1})\cr
$\sigma_{\alpha,NN'}(\omega)$
  & partial photoabsorption cross section (normalized to speed of light) 
    for basic polarization $\alpha$ ($\alpha=0,\pm1$):
    Eqs.~(\ref{sumalpha}), (\ref{GamA_NN}),  (\ref{GammA})\cr
$\sigma_\alpha^\mathrm{ff}(\omega)$, $\sigma_{\alpha,NN'}^\mathrm{ff}(\omega)$
  & total and partial cross sections (normalized to speed of light) 
    for basic polarization 
    $\alpha$, for transitions caused by free-free photoabsorption
    of a proton-electron pair:
    Eqs.~(\ref{sigmaNN'ff}), (\ref{sigma+ff}), (\ref{sigma-fit})\cr
$\sigma_{NN'}(v_z,\omega)$
  & partial cross section $\sigma_{\alpha,NN'}(\omega)$
    for longitudinal velocity $v_z$ of the absorbing particle:
    Eqs.~(\ref{muNN'}), (\ref{j})\cr
$\sigma^\mathrm{sc}_\alpha$
  & cross section of a proton with respect to scattering of a photon
    for basic polarization $\alpha$: Eq.~(\ref{sigma-sc}).
\cr}

%%%%%%%%%%%%%%%%%%%%%%%%%%%%%%%%%%%%%%%%%%%%%%%%%%%%%%%%%%%%%%%%%%%%%
\section{Order-of-magnitude estimate 
for proton Landau level populations}
\label{sect:ordermag}

In this section we present simple estimates of the relative
population of protons in the ground Landau level (number density $n_0$) 
and the first excited level ($n_1$). Other levels are neglected here for
simplicity, and also
for simplicity we assume that transitions stimulated
by radiation are unimportant. 
The cyclotron energy of the proton is $\hbar\omcp=
\hbar (eB/\mpr c)\approx 0.63 B_{14}$~keV.
In this section we use
without proof formulae for the rates of transitions
between proton Landau levels,
deferring their derivation to the following sections.

\textbf{(i) Coulomb collisions}. The collisional 
cross section involving proton
Landau transition $N=1\rightarrow N=0$ is denoted by $\sigma_{10}^{\pe}
(v_z)$, where $v_z$ is the relative velocity (along the $z$-axis) 
between the electron and the ion before collision. Detailed balance implies
$\sigma_{10}^{\pe}(v_z)=\sigma_{01}^{\pe}(v_z')$, where $v_z$ and $v_z'$ 
are related by $m_\ast v_z^2/2+\hbar\omcp=m_\ast v_z'^2/2$, and  
$m_\ast=m_e\mion/(m_e+\mion)\simeq m_e$ is the reduced mass. 
The collisional deexcitation rate per proton is 
\beq
\Gamma_{10}^{C\pe}=n_e \langle v_z\sigma_{10}(v_z)\rangle
=4\sqrt{2\pi}\,n_e\, {\am^3 e^4\sqrt{\mpr m_\ast}\over\hbar^3}
\tilde\Lambda_{10}^{\pe},
\label{Gamma10}
\eeq
where $n_e$ is the electron number density, 
$\langle\cdots\rangle$ denotes averaging over 
the 1D Maxwell distribution $f(v)\propto\exp(-m_\ast v^2/2T)$, 
and the Coulomb logarithm $\tilde\Lambda_{10}^{\pe}$ 
(to be defined later)
is of order of unity for the plasma parameters we are interested in.
In general, $\tilde\Lambda_{10}^{\pe}$ depends on parameter
$\betp \equiv \hbar\omcp/ T = 73.38\,B_{14}/T_6$,
where $T$ is the kinetic temperature for particle
motion along $\bm{B}$, and $T_6\equiv T/(10^6~{\rm K})$
(throughout this paper, we suppress Boltzmann constant,
implying the conversion 1 keV $= 1.16045\times10^7$~K).
At $\betp\gg1$ we have
$\tilde\Lambda_{10}^{\pe}\sim 1$ (and
$\tilde\Lambda_{10}^{\pe}\sim \betp^{1/2}|\ln\betp|$ for $\betp\ll 1$
(see Sect.~\ref{sect:pe}). The collisional excitation rate is 
$\Gamma_{01}^{C\pe}=\Gamma_{10}^{C\pe}\,\exp(-{\betp})$.
The contribution to Landau excitation from proton-proton collisions
is of similar order and will be neglected
     in this section.

\textbf{(ii) Radiative Transitions}. 
The spontaneous decay rate of the first Landau level is 
\beq
\Gamma^A_{10}={4\over 3}{e^2\omega_{cp}^2\over \mpr c^3}.
\label{GammA10}
\eeq 
We neglect here
the radiative absorption and stimulated emission.

\textbf{(iii) Statistical equilibrium.}
The relative population of protons in $N=0$ and $N=1$ is
determined by 
\beq
{n_1/n_0}=\mathrm{e}^{-\betp}\,
\left[ 1+\Gamma^A_{10}/\Gamma^C_{10}\right]^{-1},
\eeq

When $\Gamma^A_{10}/\Gamma^C_{10} \ll 1$, the Boltzmann distribution
(i.e., LTE) is recovered, $n_1=\mathrm{e}^{-\betp} n_0$. In this case,
the cyclotron absorption and emission are related by the 
Kirchhoff law.

In the opposite case $\Gamma^C_{10}/\Gamma^A_{10} \ll 1$, we have
$n_1/n_0=\Gamma_{01}^C/\Gamma_{10}^A$, i.e., collisional excitation
($0\rightarrow 1$) is balanced by radiative decay ($1\rightarrow 0$).

With $n_e=\rho/\mpr $, we see that
\beq
\Gamma^{C\pe}_{10}/\Gamma^A_{10}=
 28.6\,\rho_0\,B_{14}^{-7/2}\tilde\Lambda_{10}^{\pe}.
\label{GamAtoC}
\eeq
The ratio (\ref{GamAtoC})
is larger than unity for ordinary neutron star atmospheres,
but it can become smaller than unity for magnetars.

The above situation should be contrasted with that of electrons.
The deexcitation rate of
    an
 electron from its first excited
Landau level to the ground state due to collisions with 
protons (treated as classical particles) is
of the order of
$
4\sqrt{2\pi}\,\npr\, {\am^3 e^4m_e/\hbar^3}
$
 for $\beta_e
=\hbar\omce/T \gg 1$
     [cf.\ \req{Gamma10}].
The spontaneous
cyclotron decay rate of electron is 
$\Gamma^A_{10}(e)=4e^2\omce^2/(3m_ec^3)$, and the ratio
$\Gamma^C/\Gamma^A$ is about a factor $(m_e/\mpr )^{7/2}$ smaller than
that for protons. 
Thus for electrons, radiative deexcitation is always much faster than
collisional deexcitation, and there is no true
     electron
 cyclotron absorption,
but only scattering, in the magnetic fields of ordinary pulsars
and magnetars.

%%%%%%%%%%%%%%%%%%%%%%%%%%%%%%%%%%%%%%%%%%%%%%%%%%%%%%%%%%%%%%%%%%%%%
\section{Non-radiative Coulomb collision rates}
\label{sect:Coul}

Coulomb collision rates of non-degenerate fermions
in a strong magnetic field have been studied 
by many authors.
\citet{Ventura73} derived collision rates for electrons scattered
by a fixed Coulomb potential.
\citet{PavlYak} presented transition probabilities for collisions of two
non-relativistic particles, 
which interact via a screened Coulomb potential.
As a particular case they recovered the result of \citet{Ventura73},
but in a simpler form.
Relativistic expressions for Coulomb collision rates of non-degenerate fermions
in a magnetic field were derived by \citet{Langer}
and \citet{StoreyMelrose}.
However, since we are interested in Landau transitions of ions, 
we may take the non-relativistic approach \citep{PavlYak}.
Accordingly, we do not consider Coulomb spin-flip transitions
which generally are weaker by a factor $\sim b_\mathrm{i}$
compared to the transitions which preserve spin.
The spin distribution, however, may affect statistical
equilibrium through exchange effects.

%%%%%%%%%%%%%%%%%%%%%%%%%%%%%%%%%%%%%%%%%%%%%%%%%%%%%%%%%%%%%%%%%%%%%
\subsection{Proton-electron collisions}
\label{sect:pe}

General formulae for Coulomb collision rates of two different particles 
with arbitrary charges are given in Appendix~\ref{sect:z1z2}.
Here we consider electron collisions with protons,
assuming that the electrons remain in the ground Landau state.
This particular case has been previously 
considered by \citet*{MSW87}, based on \citet{PavlYak}. 
For a given relative velocity (along $z$) $v_z$ 
between a proton and an electron, the transition rate from proton
Landau level $N$ to $N'$ is
\beq
   W^\pe_{NN'} (v_z) \equiv n_e v_z \sigma_{NN'}^\pe(v_z)
   = 4\pi\tau_0^{-1} n_e\am^3 \sum_{\pm} w^\pe_{NN'}(u_\pm)/u' \!,
\label{Wpe}
\eeq
where $\sigma_{NN'}^\pe(v_z)$ is the corresponding cross section,
$\tau_0 = \hbar^3/(e^4\mel) = 2.42\times10^{-17}$~s 
is the atomic unit of time,
$n_e$ is the electron number density,
and
\beq
   w^\pe_{NN'}(u_\pm) =    
   \int_0^\infty
   \frac{\mathrm{e}^{-t/2} I_{NN'}^2(t/2) }{
   (t+ u_\pm^2 )^2} \, \dd t .
\label{wpe}
\eeq
Here $u_\pm^2 = (u\pm u')^2 + \us^2$, and
$u=(m_\ast |v_z|/\hbar)\am$ 
and $u'= (m_\ast |v_z'|/\hbar)\am$ are scaled
relative velocities along $z$,
which satisfy the energy conservation law
${u'}^2 = u^2 + 2(N-N')m_\ast/\mpr$, 
where $m_\ast=\mel\mpr/(\mel+\mpr)$ is the reduced mass.
The parameter $\us=\ks\am$, included in $u_\pm$, is the 
scaled Debye screening wave number
($\ks^{-1}$ is the Debye screening length).
For a neutral electron-proton plasma at temperature $T$ we have
$\ks=\sqrt{8\pi n_e e^2/T}=
(1.584\times 10^8\,{\rm cm}^{-1})\rho_0^{1/2}T_6^{-1/2}$.
Equations (\ref{Wpe}) and (\ref{wpe}) 
follow from (\ref{WN1N2}) and (\ref{wN1N2})
of Appendix~\ref{sect:w-general} with $Z_1=Z_2=1$ and
$w^\pe_{NN'}(u_\pm) = w_{0,N;\,0,N'}(u_\pm)$.
Laguerre function $I_{NN'}$ 
is defined by \req{Laguerre}.

If the distributions of $z$-velocities of 
electrons and protons are Maxwellian with temperatures
$T_e$ and $T_p$, respectively, which do not depend 
on the Landau number $N$,
then the relative velocities $v_z=\hbar k/m_\ast$ 
have Maxwellian distribution
\beq
   \mathcal{F}_{m_\ast,T}(v_z)=\sqrt{\frac{m_\ast}{2\pi T}}
      \,\exp\left(-\frac{m_\ast v_z^2}{2 T}\right),
\label{Maxwell}
\eeq
where 
\beq
   T = (\mel+\mpr)/(\mel T_p^{-1} + \mpr T_e^{-1}).
\label{T_ast}
\eeq
In order to simplify formulae, hereafter we assume 
$T_e=T_p=T$,
unless the opposite is explicitly stated.
Then the velocity-averaged partial Coulomb transition rate 
$n_e\langle v_z \sigma_{NN'}^\pe\rangle$ is
\begin{subequations}
\label{Gam-C}
\bea
  \Gamma^{C\pe}_{NN'} &=&
   4( e^4/\hbar^2)\,\sqrt{2\pi m_\ast/T}
      \,n_e\am^2 \Lambda^\pe_{NN'}
\label{Gam2-C}
\hspace*{3em}\\
 &=& 
 \frac{4\sqrt{2\pi}}{\tau_0}
  \left(\frac{m_\ast\mpr}{\mel^2}\right)^{\!1/2}
    n_e\am^3 \tilde\Lambda^\pe_{NN'} ,
\label{Gam1-C}
\hspace*{3em}\eea
\end{subequations}\noindent
where 
\begin{subequations}
\bea
\tilde\Lambda^\pe_{NN'} &=& \sqrt{\betp}\, \Lambda^\pe_{NN'},
\\
  \Lambda^\pe_{NN'} &=& \int_0^\infty \frac{\dd u}{u'}
   \,\mathrm{e}^{-\beta_\ast u^2/2} \,\theta({u'}^2)\, g(u)g(u')
\nonumber\\&&\times
   \left[ w^\pe_{NN'}(u_+) + w^\pe_{NN'}(u_-) \right] .
\label{Lpe}
\eea
\end{subequations}
Here $ \theta({u'}^2)$ 
[with ${u'}^2\equiv u^2+2(N-N')m_\ast/m_p$] is the step function
that ensures the energy conservation,
$ \beta_\ast \equiv {\hbar e B/m_\ast c T}
    = \betp \,\mpr/m_\ast $,
and $g(u)g(u')$ is the correction factor,
which approximately allows for violation of Born approximation
as discussed in Appendix~\ref{sect:corr}.
The latter factor appreciably differs from 1 only at $u\lesssim \gamma_B^{-1/2}$,
where $\gamma_B=\hbar^3 B/(m_\ast^2 c e^3)$.
In the case of electron-proton collisions 
$\gamma_B^{-1/2} \approx \alphaf/\sqrt{b_e}
= 0.004848\,B_{14}^{-1/2}$. 
The smallness of $\gamma_B^{-1/2}$
ensures that the approximations used to derive
\req{Lpe} are sufficiently accurate;
in this case the $\gamma_B$-dependence in \req{Lpe} is weak
     (logarithmic).

By changing integration variable $u\to u'$ in \req{Lpe},
and taking into account that $w^\pe_{NN'}(u_\pm)=w^\pe_{N'N}(u_\pm)$, 
we can check that $\Lambda^\pe_{NN'}=\mathrm{e}^{\betp(N-N')}\Lambda^\pe_{N'N}$, 
and thus
\beq
\Gamma^{C\pe}_{NN'} =\mathrm{e}^{\betp(N-N')}\Gamma^{C\pe}_{N'N}.
\eeq

\begin{figure}%
\includegraphics[width=80mm]{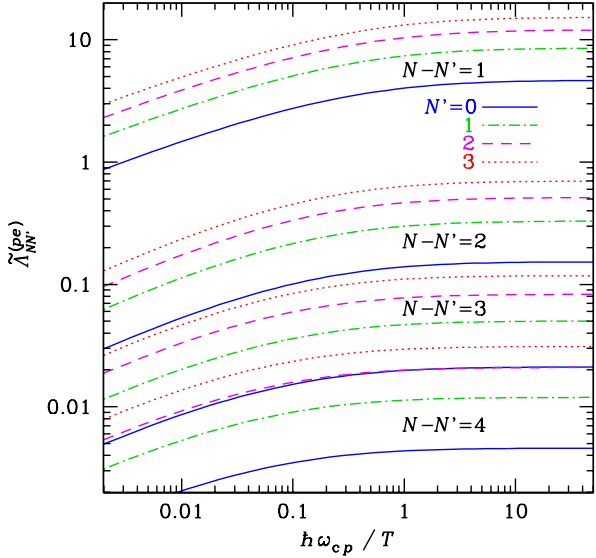}
\caption{The dimensionless quantity $\tilde\Lambda_{NN'}^\pe$ 
(Eq.~[\protect\ref{Lpe}]) as a function of $\betp=\hbar\omcp/T$
for transitions between several lowest proton Landau states
for screening parameter $\us=0$.
}
\label{fig:coule}
\end{figure}

\begin{figure}
\includegraphics[width=80mm]{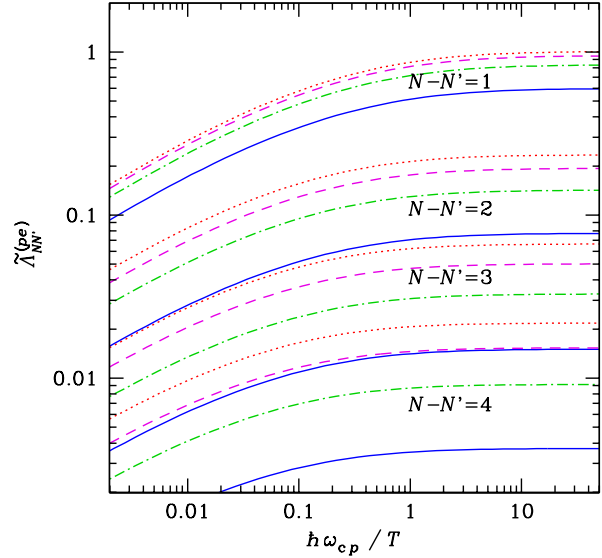}
\caption{The same as in Fig.~\ref{fig:coule},
but for screening parameter $\us=0.5$.
\label{fig:coule_s}}
\end{figure}

Figure \ref{fig:coule} depicts the 
function $\Lambda^\pe_{NN'}$ ($N>N'$)
for transitions between different low-lying proton Landau states,
calculated assuming $\us\to0$ and $\gamma_B^{-1}\to0$.
Figure \ref{fig:coule_s} presents the same functions
for $\us=0.5$. 
Note that $\us=\ks \am=4.1\times 10^{-3}
\rho_0^{1/2}T_6^{-1/2}B_{14}^{-1/2}$. So $u_s=0.5$ 
corresponds to a rather high plasma density.
We see that at any $B$ and $\us$ transitions between neighbouring states
($N-N'=1$) strongly dominate.

Representation (\ref{Gam1-C}) is most convenient when $\betp\gg1$, 
because in this case the exponential function in \req{Lpe} varies much faster than
$u'$ and $w^\pe_{NN'}$, and it can be integrated separately. Hence
$\tilde\Lambda^\pe_{NN'}$ approaches a constant.

In the opposite limit $\betp\ll1$ it may be more convenient to use
\req{Gam2-C}, because in this case
$\Lambda^\pe_{NN'}$ 
is a slowly 
varying function of $\betp$.
At the first glance it may seem unphysical that \req{Gam2-C}
contains factor $\am^2$ which goes to infinity as $B$ goes to zero.
However, it has a simple explanation. 
As long as the Landau numbers 
of the electron (equal to zero) and proton ($N,N'$) 
are kept fixed, \req{Gam2-C} describes the partial rate of the collisions in which 
the transfer of the kinetic energy of the motion transverse to the field,
$|N-N'|\hbar\omcp$, decreases
     linearly 
with decreasing $B$.
In the classical picture this corresponds to collisions 
with impact parameters increasing $\propto\am$, for which 
the cross section increases according to the Rutherford formula.
The divergence of the classical 
cross section at large impact parameters is eliminated
if one takes into account the screening of the Coulomb potential.
It is also the case for the quantum cross section.
Indeed,
$u_\pm^2$ in the denominator of \req{wpe} in general
includes the term $\us^2=\ks^2\am^2$. 
Thus $w^\pe_{NN'}(u_\pm)$
 (and hence $\Lambda^\pe_{NN'}$)
becomes $\propto(\ks\am)^{-4}$ when the magnetic length $\am$
is much larger than the screening length $\ks^{-1}$.

%%%%%%%%%%%%%%%%%%%%%%%%%%%%%%%%%%%%%%%%%%%%%%%%%%%%%%%%%%%%%%%%%%%%%
\subsection{Proton-proton collisions}
\label{sect:pp}

Now let us consider proton-proton collisions. 
This case is more complicated than the previous one 
in two respects: 
first, there is the exchange interaction 
described in Appendix~\ref{sect:like}, 
and secondly, both colliding particles can change their Landau numbers
(neither of them is confined to the ground state).

Let $n_N$ be the number density of 
protons in the $N$th Landau state,
and let $f_N^\uparrow$ and $f_N^\downarrow$ be
the fraction of
     such
protons
 with spin 
along and opposite to the field direction,
respectively ($f_N^\uparrow + f_N^\downarrow=1$). 
Then the average rate of proton transitions from level $N$
to level $N'$ due to the Coulomb collisions is
\bea
   \Gamma^{C\pp}_{NN'} &=& \frac12 {\sum_{N_2 N_2'}} n_{N_2}
   \Big[
    (f_N^\uparrow f_{N_2}^\downarrow + f_N^\downarrow f_{N_2}^\uparrow)
        \, \langle v_z \sigma_{NN_2;\,N' N_2'}^{\pp+} \rangle
\nonumber\\&&
   + (f_N^\uparrow f_{N_2}^\uparrow + f_N^\downarrow f_{N_2}^\downarrow)
        \,\langle v_z \sigma_{NN_2;\,N' N_2'}^{\pp-} \rangle
        \Big] ,
\label{Gpp}
\eea
where $\langle v_z \sigma_{NN_2;\,N' N_2'}^{\pp\pm} \rangle$ 
is the probability, per unit time, 
that two protons in unit volume, which
have initial Landau numbers $N$ and $N_2$, make a transition 
to the state where they have Landau numbers $N'$ and $N_2'$, 
under the condition 
that their spin projections to $\bm{B}$ are the same (sign $-$)
or opposite (sign $+$). The factor $\frac12$ at the sum
allows for the quantum statistics of identical particles.

For Maxwell distribution (\ref{Maxwell}) with $m_\ast=\mpr/2$,
using the results of Appendix \ref{sect:2part},
we obtain
\bea&&\hspace*{-2em}
  \langle v_z \sigma_{NN_2;\,N' N_2'}^{\pp\pm} \rangle
   = 8\, (e^4/\hbar^2)
      \,\sqrt{\pi\mpr/T} \,\am^2
\nonumber\\&&\times
      \int_0^\infty \frac{\dd u}{u'}\,\mathrm{e}^{-\betp u^2}\,\Big[
         w_{N N_2;\,N' N_2'}(u_+) + w_{N N_2;\,N' N_2'}(u_-)
\nonumber\\&&\hspace*{2em}
          \pm
       2\, w_{N N_2;\,N' N_2'}^x(u_-,u_+)
         \Big] ,
\label{gam-pp}
\hspace*{3em}\eea
where
$u_\pm^2 = (u\pm u')^2 + \us^2$
and $u' = [u^2 + N' - N + N_2' - N_2]^{1/2}$.
Functions $w_{N N_2;\,N' N_2'}$ and $w_{N N_2;\,N' N_2'}^x$
are given by equations (\ref{wN1N2}) and (\ref{wx}), respectively.

Equations (\ref{Gpp}) and (\ref{gam-pp}) can be written 
in the form analogous to \req{Gam1-C},
\bea&&\hspace*{-2em}
   \Gamma^{C\pp}_{NN'} = \sum_{N_2}
       n_{N_2} \langle v_z \sigma_{NN'}^{(N_2)} \rangle ,
\\&&\hspace*{-2em}
   \langle v_z \sigma_{NN'}^{(N_2)} \rangle =
   4\sqrt{\pi} \, \frac{\am^3}{\tau_0}\,
   \frac{\mpr}{\mel}
   \, \Big[ \tilde\Lambda_{N N_2;\,N'}^\pp
\nonumber\\&&
   - (f_{N}^\uparrow - f_N^\downarrow)
   \,(f_{N_2}^\uparrow - f_{N_2}^\downarrow)
   \,\tilde\Lambda_{N N_2;\,N'}^{\ppx}
   \Big] ,
\label{GppL}
\eea
where
\begin{subequations}
\label{Lam-pp}
\bea&&\hspace*{-2em}
   \tilde\Lambda_{N N_2;\,N'}^\pp
   = \sqrt{\betp} \sum_{N_2'} \int_0^\infty
    \frac{\dd u}{u'}\,\mathrm{e}^{-\betp\,u^2}\,\theta({u'}^2)\, g(u)g(u')\,
\nonumber\\&&\times
    [ w_{N N_2;\,N' N_2'}(u_+) + w_{N N_2;\,N' N_2'}(u_-) ],
\label{Lpp}
\hspace*{3em}\\&&\hspace*{-2em}
   \tilde\Lambda_{N N_2;\,N'}^{\ppx}
   = 2\,\sqrt{\betp} \sum_{N_2'} \int_0^\infty
 \frac{\dd u}{u'}\,\mathrm{e}^{-\betp\,u^2}\,\theta({u'}^2)\, g(u)g(u')\,
\nonumber\\&&\times
  w_{N N_2;\,N' N_2'}^x(u_-,u_+) .
\eea
\end{subequations}\noindent
In these equations $u'$ and
$u_\pm$ depend on $N_2'$. 
The factors $g(u)g(u')$, with $g(u)$ defined
by \req{gBorn}, account for the 
correction due to violation of Born approximation, as 
discussed in 
Appendix~\ref{sect:corr}.
However, unlike Sect.~\ref{sect:pe}, here
$\gamma_B^{-1/2} = \alphaf\mpr/(2\mel\sqrt{b_e}) 
= 4.45\,B_{14}^{-1/2}$
is larger than 1 for $B < 2\times10^{15}$~G,
which reflects the fact that the protons 
are moving much slower than the electrons,
therefore Born and adiabatic approximations
(see Appendix~\ref{sect:corr}) are less applicable to 
the proton-proton collisions. Nevertheless, we use these approximations,
considering them as order-of-magnitude estimates,
which is justified
because the whole effect of the proton-proton collisions 
on statistical equilibrium is not very significant,
as we will see below. 

\begin{figure}
\includegraphics[width=80mm]{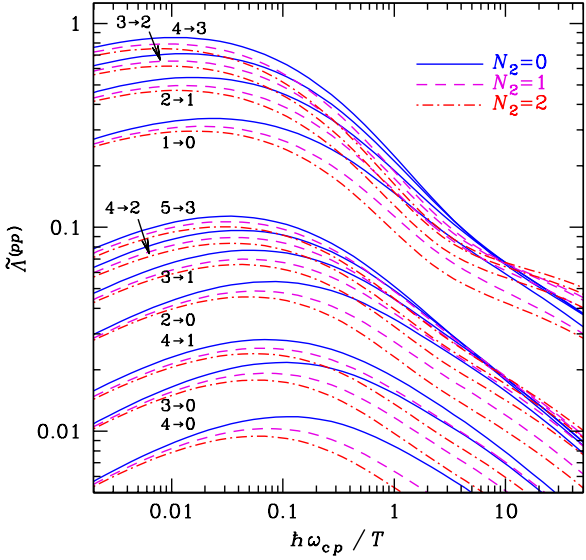}
\caption{
The dimensionless function $\tilde\Lambda_{NN_2;\,N'}^\pp$ 
(Eq.~[\ref{Lpp}])
without screening ($\us=0$) for the transitions 
between the proton Landau states marked near the curves.
Initial Landau number of the second proton equals
$N_2=0$ (solid lines), $N_2=1$ 
(dashed lines), or $N_2=2$ (dot-dashed lines).
\label{fig:coulpp}}
\end{figure}

\begin{figure}
\includegraphics[width=80mm]{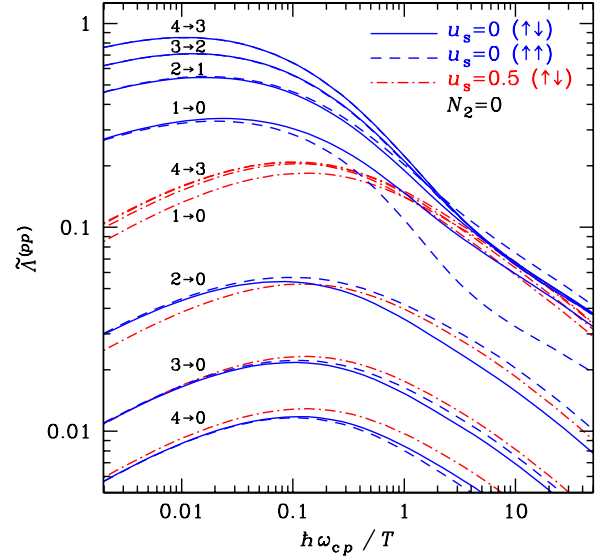}
\caption{The functions $\tilde\Lambda_{NN_2;\,N'}^\pp$
(solid lines without screening and dot-dashed lines
with screening parameter $\us=0.5$)
and $\tilde\Lambda_{NN_2;\,N'}^\pp- \tilde\Lambda_{NN_2;\,N'}^{\ppx}$
(dashed lines, without screening), representing 
the expression in the square brackets of \req{GppL}
for random spin orientations ($f^\uparrow=f^\downarrow=0.5$)
and for aligned spins ($f^\uparrow=1$),
respectively, for transitions $N\to N-1$ and $N\to0$,
     with
$N=1,2,3,4$, $N_2=0$,
as functions of $\betp=\hbar\omcp/T_p$.
\label{fig:coulpp_s}}
\end{figure}

In Figure \ref{fig:coulpp} we show $\tilde\Lambda_{NN_2;\,N'}^\pp$
for the case of negligible screening ($\us=0$) 
and $\gamma_B=1$,
for a few initial and final
proton Landau numbers $N,N'$, and for 
initial Landau numbers of the second proton $N_2=0,1,2$. 
The decrease of the displayed functions at $\betp\gg1$ 
results from the correction beyond Born approximation.
This indicates that an accurate evaluation 
of the proton-proton collision rates would require non-Born
quantum calculations, which are beyond the scope of the present paper.

In figure \ref{fig:coulpp_s} 
we compare some of the curves from figure \ref{fig:coulpp}
(solid lines)
with the case of non-negligible screening, $\ks\am=\us=0.5$
(dot-dashed lines),
which can be relevant at rather high density.
Also in this figure we show a comparison of 
$\tilde\Lambda_{NN_2;\,N'}^\pp$ (solid lines) with 
the difference 
$\tilde\Lambda_{NN_2;\,N'}^\pp- \tilde\Lambda_{NN_2;\,N'}^{\ppx}$
(dashed lines)
which enters \req{GppL} when the proton spins are all aligned 
in the same direction.

Let us note that if $T_e\neq T_p$, then
in this section $T_p$ should substitute $T$ defined by \req{T_ast}.
In particular, $\betp=\hbar\omcp/T_p$ in 
Eqs.\ (\ref{gam-pp}) and (\ref{Lam-pp}).

%%%%%%%%%%%%%%%%%%%%%%%%%%%%%%%%%%%%%%%%%%%%%%%%%%%%%%%%%%%%%%%%%%%%%
\section{Radiative transitions}
\label{sect:Rad}

%%%%%%%%%%%%%%%%%%%%%%%%%%%%%%%%%%%%%%%%%%%%%%%%%%%%%%%%%%%%%%%%%%%%%
\subsection{Radiative transition rates in magnetized plasmas}
\label{sect:RadRates}

Magnetized plasma is a birefringent medium.
Electromagnetic radiation propagates through it
in the form of two normal polarization modes $j=1,2$ with polarization
vectors $\bm{e}^j(\omega,\dir)$
(e.g., \citealt{Ginzburg}).
Here, $\omega$ is the angular frequency and $\dir$
the unit vector along the wave vector.
Consequently, radiative transition rates depend not only on $\omega$,
but also on $j$ and $\dir$.

Let $\Gamma^A_{NN'}$, $\Gamma^B_{NN'}$, and $\hat\Gamma^B_{NN'}$ be
the rates of transitions from level $N$ to $N'$ due to spontaneous
emission, photoabsorption, and stimulated emission, respectively.%
     \footnote{Generally, $N$ and $N'$ may take any values.
     For instance, free-free photoabsorption is allowed
     for $N'\leq N$ 
(i.e., the photon is absorbed while the proton makes a downward transition)
as well as for $N'>N$.}
These rates are the \emph{total} (for both polarizations, integrated
over angles and frequencies) transition probabilities per unit time
for one occupied initial quantum state.  They can be expressed through
Einstein coefficients $A_{NN'}$ and $\hat{B}_{NN'}$ (for emission), or
$B_{NN'}$ (for absorption).  These coefficients have different
definitions in the literature (e.g., cf.\ \citealt{RybickiLightman},
\citealt{Ginzburg}, and \citealt{Zhel}).  We define $A_{NN'}$,
$B_{NN'}$, and $\hat{B}_{NN'}$ from the conditions that the number of
quanta with angular frequencies in the interval $\dd\omega$ and wave
vectors in solid angle element $\dd\dir$ spontaneously emitted by a
unit volume during unit time equals $n_N A_{NN'}\,\dd\omega\,\dd\dir$,
and the number of quanta emitted or absorbed under the action of
radiation with the specific intensity $I_\omega$ equals $n_N
\hat{B}_{NN'} I_\omega\,\dd\omega\dd\dir$ or $n_N B_{NN'}
I_\omega\,\dd\omega\dd\dir$, respectively \citep{Zhel}.  This
definition (or a similar one in \citealt{Ginzburg}, but not the one in
\citealt{RybickiLightman}) is relevant in a strong magnetic field,
where the emission is neither isotropic, nor unpolarized.  Then
\begin{eqnarray}
   \Gamma^B_{NN'} &=& \sum_{j=1,2} \int\dd\dir \int\dd\omega
      B_{NN'}(j,\omega, \dir)
      \, I_{\omega,j}(\dir),
\label{Gam-B}
\\
   \Gamma^A_{NN'} &=& \sum_{j=1,2} \int\dd\dir \int\dd\omega
            \, A_{NN'}(j,\omega, \dir) ,
\label{Gam-A}
\end{eqnarray}
and the expression for $\hat\Gamma^B_{NN'}$ is completely analogous
to \req{Gam-B}.

The quantities $A_{NN'}$, $\hat{B}_{NN'}$ and $B_{N'N}$ are related by 
the Einstein relations
(which include polarization dependence, see, e.g.,
\citealt{Ginzburg,Zhel}):\footnote{
Einstein relations (\ref{Ein-2})
come from the detailed balance equation
$n_{N'} B_{N'N} I_{\omega,j}  = n_N A_{NN'} + n_N \hat{B}_{NN'} I_{\omega,j}$
in the complete thermodynamic equilibrium, 
where $I_{\omega,j}=\frac12 B_\omega = 
(\hbar\omega^3/8\pi^3c^2)\,(\mathrm{e}^{\hbar\omega/T}-1)^{-1}$,
and the requirement that the coefficients $A$ and $B$ must
be independent of $T$.
}
\beq
  \hat{B}_{NN'} = B_{N'N},
\quad
   A_{NN'} = \frac{ \hbar\omega^3 }{ 
   8\pi^3 c^2 }
   \, B_{N'N} .
 \label{Ein-2}
\eeq
From the first of these relations, it follows that the
stimulated emission rate is equal to that of
photoabsorption with interchange of the initial
and final levels: $\hat\Gamma^B_{NN'}=\Gamma^B_{N'N}$.

In \req{Ein-2} we have neglected the difference of the group and phase
velocities of radiation. 
Einstein relations with allowance for this difference are given,
e.g., by \citet{Ginzburg} and \citet{Zhel}.
Note, however, that this difference would lead to appearance
of additional factors not only in \req{Ein-2}, but also in
the expressions for photoabsorption cross sections discussed 
in Sect.~\ref{sect:Gamma-sigma} below. 

Spontaneous cyclotron decay rates have been derived by \citet{DV77} 
(see also \citealt*{DV78,MZh,Pavlov-ea91,Baring-ea05}, 
and references therein).
In the non-relativistic limit ($N b_\mathrm{i} \ll 1$),
the decay rates are proportional to $b_\mathrm{i}^{N-N'+1}$,
multiplied by a combinatorial factor. Although the latter factor
may be large for large 
$N'-N$, 
transitions $N\to N'=N-1$ still dominate 
in the non-relativistic regime. 
The rates of the latter transitions are
$
   \Gamma^A_{N,N-1} = N\Gamrp,
$
where
\beq
   \Gamrp=\frac43\,\frac{e^2\omcp^2}{\mpr\, c^3} 
\eeq
is the natural width of 
the proton cyclotron line.

Spin-flip transitions 
for protons
are unimportant in the non-relativistic limit, because their
rates contain an additional factor $b_\mathrm{i}$ compared to 
the dominant transitions preserving spin
(e.g., cf.\ \citealt{MZh}).

%%%%%%%%%%%%%%%%%%%%%%%%%%%%%%%%%%%%%%%%%%%%%%%%%%%%%%%%%%%%%%%%%%%%%
\subsection{Relation between emission rates and
 photoabsorption cross sections}
\label{sect:Gamma-sigma}

The Einstein absorption coefficient $B_{NN'}$ is given by the relation
\beq
     B_{NN'}(j,\omega, \dir) = \sigma_{NN'}(j,\omega,\dir)/\hbar\omega ,
\label{BbarJ}
\eeq
where $\sigma_{NN'}$ is the partial photoabsorption 
cross section responsible for the $N\to N'$ transition.
Equation (\ref{BbarJ}) directly follows from the definition of
$B_{NN'}$ in Sect.~\ref{sect:RadRates}. Together with Einstein relations,
it allows one to express spontaneous decay rates
$\Gamma^A_{NN'}$ through 
partial photoabsorption cross sections.

In the `rotating coordinates' \citep[e.g.][]{Mesz},
the polarization vectors $\bm{e}^j(\omega,\dir)$ 
of two polarization modes $j=1,2$ have the components 
$e_\alpha^j$, $\alpha=0,\pm1$.
In the dipole approximation,
photoabsorption
cross sections can be written as
(e.g., \citealt*{Ventura-ea79})
\beq
   \sigma(j,\omega,\dir) = 
     \sum_{\alpha=-1}^1 \sigma_\alpha\low(\omega)
      \,\,|e_\alpha^j(\omega,\dir)|^2,
\label{sumalpha}
\eeq
where 
the component $e_-$ is responsible 
for the electron cyclotron resonance,
and $e_+$ for the ion cyclotron resonance. 

Using equations (\ref{Gam-A}), (\ref{Ein-2}) and (\ref{BbarJ}),
one obtains
\beq
   \Gamma^A_{N'N} = \sum_{j=1,2} \int{\dd\dir}
  \int \frac{\omega^2 \dd\omega}{8\pi^3 c^2 }
  \, \sigma_{NN'}(j,\omega,\dir)\,.
\label{GamA_NN}
\eeq

As already stated in Sect.~\ref{sect:RadRates}, we neglect
the difference of the group and phase
velocities of radiation. This is equivalent to the
`semi-transverse 
approximation'
\citep{Ventura79}, where refraction indices are close to 1,
and $\dir\cdot\bm{e}^j\approx0$.
In this approximation, the relation
$\sum_{j=1,2} A_\alpha^j = 1$ holds,
where
\beq
   A_\alpha^j \equiv \frac{3}{8\pi}\int|e_\alpha^j(\omega,\dir)|^2\,\dd\dir\,.
 \label{A}
\eeq
Then equations (\ref{sumalpha}) and (\ref{GamA_NN})
give
\beq
  \Gamma^A_{N'N} =  \sum_\alpha
  \int \frac{\omega^2\,\dd\omega}{3\pi^2\,c^2 }
  \, \sigma_{\alpha,NN'}(\omega).
\label{GammA}
\eeq

Let us suppose that
transition $N\to N'$ corresponds to an absorption line
with a profile $\phi(\omega)$ 
($\int\phi(\omega)\dd\omega=1$), which
at $\omega=\omega_0$ has a sharp peak with a characteristic
     half-width $\nu\ll\omega_0$
for the polarization $\alpha$.
At $\omega\sim\omega_0$, let us write the photoabsorption cross section
in the form $\sigma_{\alpha,NN'}(\omega)=
    2\nu
\bar\sigma_{\alpha,NN'}\phi(\omega)$.
Then \req{GammA} gives the spontaneous emission rate
\beq
\Gamma^A_{N'N}\approx
       \frac{2\nu\bar\sigma_{\alpha,NN'}\,\omega_0^2}{
             3\pi^2c^2}.
\eeq

Let us assume, in addition, that $I_{\omega,j}(\dir)$ 
can be replaced by the average over the angles under 
the integral in \req{Gam-B}, which we denote
$\bar{I}_{\omega,j}$
(this replacement is exact in the diffusion approximation).
Then, 
using equations (\ref{BbarJ}) and (\ref{A}),
we obtain the photoabsorption and stimulated emission rates
\beq
   \Gamma^B_{NN'} 
   = \hat\Gamma^B_{N'N} =
  \frac{8\pi^3c^2}{\hbar\omega_0^3} \,\Gamma^A_{N'N}
   \sum_{j=1,2} A_\alpha^j\,J_j  ,
\label{Gamma01}
\eeq
where
\beq
  J_j \equiv \int \, \bar{I}_{\omega,j}\,\phi_j(\omega)\,\dd\omega .
\label{barJ}
\eeq

%%%%%%%%%%%%%%%%%%%%%%%%%%%%%%%%%%%%%%%%%%%%%%%%%%%%%%%%%%%%%%%%%%%%%
\subsection{Cross sections at the proton cyclotron resonance}
\label{sect:cyclo}

Although cyclotron emission rates can be calculated
in the framework of the first-order perturbation theory,
this theory is not suitable for the
determination of the frequency dependence of photoabsorption
cross sections and opacities. The reasons for that,
and the conditions where the first-order process 
still can be important, were discussed by \citet{DV78},
who stressed that the spectral dependence of the
absorption coefficient is properly described
by second- and higher-order 
processes.
Indeed, because of kinematic requirements (energy and momentum conservation),
 the first-order absorption
is possible only at a single frequency
at any given angle of incidence.
Thus one must take into account level broadening
in order to obtain the spectral absorption coefficient.
The broadening is caused by the finite life time of
the proton in a final state after an absorption event.
In a macroscopically homogeneous plasma
this life time is limited only by (1) spontaneous emission
and (2) interactions
with other particles. It is the emitted photon in the first case
or another plasma particle in the second case that carries
the energy and momentum needed to restore the kinematic balance.
Thus a quantum description of the absorption line shape 
requires at least two-vertex Feynman diagrams.
The second vertex
may correspond to the emission of the photon [case (1);
cf.\ Figs.~1 and 2 of \citealt{DV78}]
or to Coulomb interaction with a charged particle [case (2)].
The first case is scattering, and
the second is free-free photoabsorption.
We now consider the
cross sections for these two processes
for the polarization component $\alpha=+1$,
corresponding to the proton cyclotron resonance.

%%%%%%%%%%%%%%%%%%%%%%%%%%%%%%%%%%%%%%%%%%%%%%%%%%%%%%%%%%%%%%%%%%%%%
\subsubsection{Free-free absorption}

The free-free photoabsorption cross section 
at any frequencies and polarizations is given by
     Eqs.~(\ref{sigma-fit})\,--\,(\ref{damp})
of Appendix~\ref{sect:ff}.
It can be presented as
a sum of terms corresponding to transitions of a proton
from level $N$ to $N'$:
\beq
\sigma_\alpha^\mathrm{ff}(\omega) = \sum_N f^p_N \sum_{N'}
   \sigma_{\alpha,NN'}^\mathrm{ff}(\omega),
\label{sigma-sep}
\eeq
     where
     $f^p_N$ is the fraction of protons in Landau state $N$,
      and  
\beq
   \sigma_{\alpha,NN'}^\mathrm{ff}(\omega) = 
     \frac{4\pi e^2}{\mel c}
   \frac{\omega^2
   \sum_n f^e_n \sum_{n'}\nu_{n,N;\,n',N'}^{\mathrm{ff},\alpha}(\omega)
          }{
          (\omega+\alpha\omce)^2 (\omega-\alpha\omcp)^2
             +\omega^2 \tilde\nu_\alpha^2(\omega)},
\label{sigmaNN'ff}
\eeq
     where $f^e_n$ is the fraction of the electrons in Landau state $n$,
     and $\tilde\nu_\alpha(\omega)$ is a damping factor.
The separation of $\sigma_\alpha^\mathrm{ff}$ into $\sigma_{\alpha,NN'}^\mathrm{ff}$,
expressed by \req{sigma-sep},
is useful for calculation of non-LTE emissivity
(see Sect.~\ref{sect:opem}).

At $\omega\sim\omcp$ and $\alpha=+1$, there is a resonance:
\beq
   \sigma_{+}^\mathrm{ff}(\omega) \approx
   \frac{4\pi e^2}{\mpr c}\,
   \frac{\nu_p^\mathrm{ff}}{
   (\omega-\omcp)^2+\nu^2} ,
\label{sigma+ff}
\eeq
where $\nu=
(m_e/m_p)\tilde\nu_+(\omcp)=
\hat\nu_p+ \nu_p^\mathrm{ff}$, with $\hat\nu_p=\nu_p(\omcp)$
the radiative damping rate $\Gamma_{r,p}/2$
(note that it could also include other damping mechanisms 
not related to electron-proton collisions, such as the damping rate 
due to collisions with neutral particles), and
$\nu_p^\mathrm{ff}$ the damping rate due to electron-proton collisions:
\beq
  \nu_p^\mathrm{ff}=(\mel/\mpr)\,\nu_{+}^\mathrm{ff}(\omcp),
\label{nup}
\eeq
with
$\nu_{+}^{\mathrm{ff}}(\omega)$ given by \req{nu-ff}.
From \req{GammA}, taking into account
the condition $\nu\ll\omcp$,
we obtain the rate of
$N'\to N$ transitions caused by 
the resonant free-free emission,
\bea&&\hspace*{-2em}
   \Gamma^{\mathrm{ff}}_{+,N'N} \approx
     \frac{4\sqrt{2\pi}}{\tau_0}
       \left(\frac{\mpr}{\mel}\right)^{\!1/2} n_e \am^3\,
\nonumber\\&&\times
   \frac{\sqrt{\betp}}{3\pi} \int  % \frac{\omega}{\omcp}
   \,\frac{\Gamrp
  \,\Lambda_{0,N;\,0,N'}^{\mathrm{ff},+1}  % (\beta_\ast,\omega/\omega_\ast)
}{
   (\omega-\omcp)^2+\nu^2}
   \, \dd \omega ,
\label{GammAff}
\hspace*{2em}\eea
where 
$\Lambda_{n,N;\,n',N'}^{\mathrm{ff},+1} % (\beta_\ast,\omega/\omega_\ast)
$
is given by \req{Lambda1}.
This result is written
in the form similar to \req{Gam1-C} for easy comparison,
which shows that by order of magnitude
$\Gamma^{\mathrm{ff}}/\Gamma^{C\pe}
 \sim \Gamrp/\nu$.
     The damping factor $\nu$ is discussed in Sect.~\ref{sect:sc};
     here we note only that the ratio
     $\Gamma^{\mathrm{ff}}/\Gamma^{C\pe}$  cannot be large,
     because $\nu\geq\Gamrp/2$.

%%%%%%%%%%%%%%%%%%%%%%%%%%%%%%%%%%%%%%%%%%%%%%%%%%%%%%%%%%%%%%%%%%%%%
\subsubsection{Scattering}
\label{sect:sc}

The resonant cyclotron scattering \citep*{Canuto-ea,Ventura79}
is a second-order process
which is common for electrons in white dwarfs and magnetic neutron stars,
and for ions in magnetars.
The photon-proton scattering cross section 
is 
\beq
   \sigma^\mathrm{sc}_{+} = \sigma_{\mathrm{T}p}
   \,\frac{\omega^2}{(\omega-\omcp)^2 + \nu^2} ,
\label{sigma-sc}
\eeq
where $\sigma_{\mathrm{T}p}=8\pi e^4/(3\mpr^2 c^4)$
is the Thomson cross section for protons.

The determination of the effective damping factor $\nu$
(not considered by \citealt{Canuto-ea}) is not trivial.
In general, this task requires a non-perturbative treatment
\citep*{Cohen-T-ea}, which goes beyond the scope of our paper.
However, $\nu(\omega)$ can be found from the correspondence 
to the classical physics.

The na\"{\i}ve estimate
     of $\nu$ as the sum of total half-widths of two Landau levels
would lead to replacement of \req{sigma-sc} by a sum
of different Lorentz profiles for different proton states $N$.
However, this estimate is incorrect,
because it ignores the coherence of equally spaced quantum states,
as discussed, e.g., by \citet{Cohen-T-ea}
for the case of interaction of electromagnetic field with a quantum
oscillator. Interference 
of transition amplitudes between different states
leads to the common damping factor 
(which proves to be equal 
to the classical oscillator damping factor)
for all transitions
which have the same resonant frequency.
Thus
we should put in \req{sigma-sc}
the same damping factor as in \req{nup},
$\nu=\hat\nu_p+\nu_p^\mathrm{ff}$ at $\omega\approx\omcp$.
The frequency dependence of $\nu$ is suggested 
by analogy with a classical oscillator \citep{Jackson}:
\beq
   \nu(\omega)=\nu_p(\omega)+(\mel/\mpr)\,\nu_+^\mathrm{ff}(\omega),
\eeq
with
\beq
   \nu_p(\omega) = \frac23\,\frac{e^2}{\mpr c^3}\,\omega^2.
\eeq
Thus we recover the damping factor that was previously given
without discussion by \citet{Pavlov95}.
Obviously, at the resonance, $\nu_p(\omcp)=\hat\nu_p=\Gamrp/2$.

Note that for damping of free-free photoabsorption (Eq.~[\ref{damp}])
we should include, beside $\nu_p(\omega)$,
also $\nu_e(\omega)=(2/3)(e^2/\mel c^3)\,\omega^2$.
Then the terms containing factor $\alpha$ in \req{damp} 
cancel out, and it simplifies to
\beq
  \tilde\nu_\alpha(\omega)  = \frac23\,\frac{e^2}{m_\ast c^3}\,\omega^2
    + \nu_\alpha^\mathrm{ff}(\omega).
\label{damp-simple}
\eeq

%%%%%%%%%%%%%%%%%%%%%%%%%%%%%%%%%%%%%%%%%%%%%%%%%%%%%%%%%%%%%%%%%%%%%
\section{Statistical equilibrium of proton Landau levels}
\label{sect:stateq}

%%%%%%%%%%%%%%%%%%%%%%%%%%%%%%%%%%%%%%%%%%%%%%%%%%%%%%%%%%%%%%%%%%%%%
\subsection{Two-level system}
\label{sect:2level}

The model in which only two quantum levels participate in the radiative
and collisional transitions is helpful for 
understanding
the main features of
line formation and transition rates. This simplest model can be 
applicable to the formation of 
the proton cyclotron line if 
the ground Landau level 
is much more populated than excited ones.
For \textit{electron} cyclotron lines, 
a similar model was considered previously by \citet{NagelVentura}.

The statistical equilibrium of two proton Landau levels is given by the 
equation
\beq
   n_0 \left( \Gamma^B_{01} + \Gamma^C_{01} \right)
  = n_1 \left( \Gamma^A_{10} + \hat\Gamma^B_{10} + \Gamma^C_{10} \right).
\label{steq10}
\eeq
Let us first consider 
the case where proton-proton collisions are unimportant.
Then, taking into account \req{Gamma01} and the relation
$\Gamma^{C\pe}_{01} = \Gamma^{C\pe}_{10} \mathrm{e}^{-\betp}$ 
(Sect.~\ref{sect:pe}), 
\req{steq10}
can be written in the form
\beq
   \frac{n_1}{n_0} = \mathrm{e}^{-\betp} \,
   \frac{ 1 + \epsilon\,R/(1-\mathrm{e}^{-\betp}) }{
      1+ R + \epsilon\,R/(\mathrm{e}^\betp-1) } ,
      \quad
      R\equiv\Gamrp/\Gamma^C_{10}
\label{balance10}
\eeq
where the parameter
$
   \epsilon = 2\sum_j A_+^j J_j / B_\omega
$
(at $\omega=\omcp$)
characterizes the ratio of the effective radiative energy density in the line
to its equilibrium value.

Let us mention three important limiting cases.
\begin{itemize}
\item
When either $R\ll1$ or $\epsilon=1$,
the Boltzmann ratio $n_1/n_0=\mathrm{e}^{-\betp}$ is recovered. 
This is the LTE situation, where absorption and emission coefficients
are related by the Kirchhoff law.
\item
In another limiting case,
where $R\gg1$ and $\epsilon R\ll1$, 
$n_1/n_0 = \mathrm{e}^{-\betp}/R = \Gamma^C_{01}/\Gamrp$, that is, 
excitation of the level $N=1$ is collisional,
but its deexcitation is radiative.
This is the case for which emission is most prominent.
\item
In the third limit, where $\epsilon\ll1$ and $\epsilon R\gg1$,
the level $N=1$ is excited by absorption of radiation
and deexcited by spontaneous emission. In this case
$n_1/n_0 = (8\pi^3 c^2/\hbar\omega^3)\sum_j A_+^j J_j$,
and $$n_1 \sum_j \int\dd\dir A_{10}(j,\omega,\dir)
= n_0 \sum_j \int\dd\dir B_{01}(j,\omega,\dir) \bar{I}_{\omega,j}.$$
Then the spectral power of spontaneous
emission is identical to that of absorption, and both processes
can be treated as non-coherent scattering \citep{Ventura79,Mesz}.
Such situation is most common for the electron cyclotron
absorption and emission in strong magnetic fields
of neutron stars
\citep{NagelVentura}, but 
it is not so usual for
     ion (proton)
 cyclotron processes.
\end{itemize}

Taking into account proton-proton collisions, % (Sect.~\ref{sect:pp}),
from \req{steq10} we obtain
\bea&&\hspace*{-2em}
   \frac{n_1}{n_0} = \frac{1}{2+2\,c_{10}^{(1)}}
   \Big\{ \big[(1+c_{10}^{(0)}-c_{01}^{(1)}-x_1)^2
\nonumber\\&&\hspace*{-1em}
      +4\,(1+c_{10}^{(1)})\,(x_1+c_{01}^{(0)})\big]^{1/2}\!
      -(1+c_{10}^{(0)}-c_{01}^{(1)}-x_1) \Big\},
\label{balance10c}
\eea
where $c_{NN'}^{(N_2)} = \frac12 
\npr \langle v_z \sigma_{NN'}^{(N_2)} \rangle
 /
(\Gamma^A_{10}+\Gamma^B_{01}+\Gamma^{C\pe}_{10})$,
the factor
$\langle v_z \sigma_{NN'}^{(N_2)} \rangle$ is given by \req{GppL},
and $x_1$ is the solution (\ref{balance10}), which is reproduced
when $c_{NN'}^{(0,1)}\to0$.

\begin{figure}
\includegraphics[width=80mm]{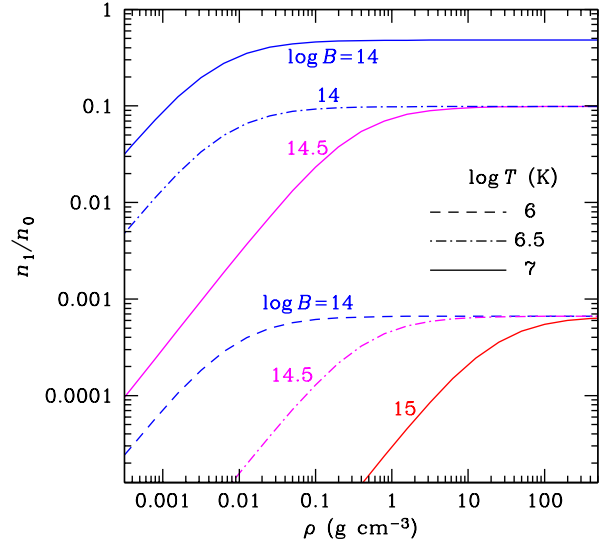}
\caption{Population of the 
proton Landau 
level $N=1$ relative to $N=0$
as function of mass density 
for different values of $B$ and $T$.
\label{fig:steq2}}
\end{figure}

In Figure \ref{fig:steq2} we show the relative populations
of the excited proton Landau state ($N=1$) as function of density, according
to \req{balance10c}, for several $B$ and $T$ values.
Here we assumed that stimulated transitions are unimportant
and set $\epsilon=0$ in \req{balance10}.
At high density, all curves tend to their LTE
limit $n_1/n_0=\mathrm{e}^{-\betp}$. Radiative decay rates
dominate at lower densities,
where the excited level becomes depopulated.

For the considered plasma parameters, the rates 
of transitions due to proton-proton collisions 
are of the same order of magnitude or smaller
than those due to electron-proton collisions.
Therefore a neglect of the $pp$ rates does not significantly
affect the statistical equilibrium.
For instance, in Fig.~\ref{fig:steq2} this neglect would change
$n_1/n_0$ by less than 6 per cent.
Thus statistical equilibrium can be approximately
evaluated with only proton-electron interactions taken into account.

%%%%%%%%%%%%%%%%%%%%%%%%%%%%%%%%%%%%%%%%%%%%%%%%%%%%%%%%%%%%%%%%%%%%%
\subsection{Multilevel system}
\label{sect:multilevel}

The statistical equilibrium of proton distribution over Landau levels
is determined by the balance of the total rates of transitions
from and to every level $N$,
\bea&&\hspace*{-2em}
   n_N\low \Bigg[ \sum_{N' < N} \Gamma^A_{NN'}
     + \sum_{N'\neq N} ( \Gamma^B_{NN'} + \Gamma^C_{NN'}) \Bigg]
     =
     \sum_{N' > N} n_{N'}\low \,\Gamma^A_{N'N}
\nonumber\\&&
  + \sum_{N'\neq N} n_{N'}\low\,( \Gamma^B_{N'N} + \Gamma^C_{N'N}) ,
\label{stateq}
\eea
supplemented with the condition $\sum_N n_N = \npr$.
Here $\Gamma^C_{NN'}=\Gamma^{C\pe}_{NN'}+\Gamma^{C\pp}_{NN'}$.
This system is non-linear, because 
according to \req{Gpp}
$\Gamma^{C\pp}_{NN'}$ depends on the distribution of $n_N$. 
We solve \req{stateq} iteratively, starting from
the Boltzmann distribution of $n_N$.

\begin{figure}
\includegraphics[width=80mm]{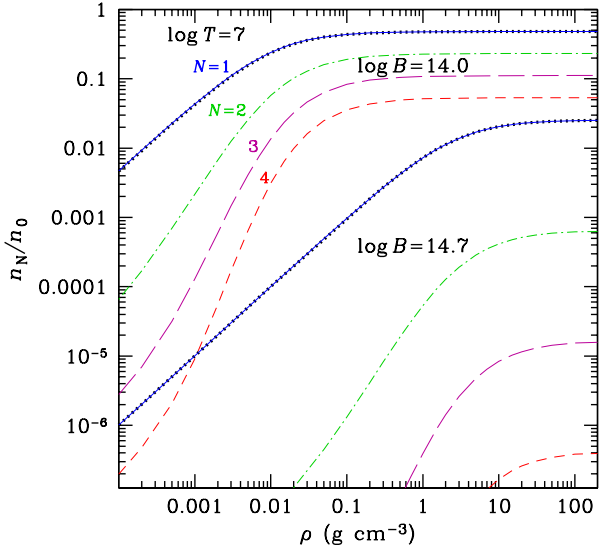}
\caption{Populations of the levels $N=1,2,3,4$ relative to $N=0$
as functions of mass density 
for $T=10^7$~K and two values of $B$. 
The solid lines show the population of the $N=1$ level,
dot-dashed lines $N=2$, long-dashed lines $N=3$, and
short-dashed lines $N=4$. The dotted lines show $n_1/n_0$ based on 
the two-level approximation.
\label{fig:steq}}
\end{figure}

An example of the numerical solution of \req{stateq} is shown 
in Fig.~\ref{fig:steq}.
As in Fig.~\ref{fig:steq2}, here we have neglected $\Gamma^B$
relative to $\Gamma^C$.
The curves of different style correspond to the values
of $n_N/n_0$ as functions of $\rho$ for different $N$
from 1 to 4. The dots correspond to $n_1/n_0$
in the two-level approximation (cf.\ Fig.~\ref{fig:steq2}).
We see that they coincide with the multilevel solution
for $n_1/n_0$ (solid lines) within graphical accuracy.

%%%%%%%%%%%%%%%%%%%%%%%%%%%%%%%%%%%%%%%%%%%%%%%%%%%%%%%%%%%%%%%%%%%%%
\section{Opacity and emissivity}
\label{sect:opem}

%%%%%%%%%%%%%%%%%%%%%%%%%%%%%%%%%%%%%%%%%%%%%%%%%%%%%%%%%%%%%%%%%%%%%
\subsection{Relation between emission and absorption coefficients}

For each polarization component, the 
photoabsorption coefficient can be presented in the form
\bea&&\hspace*{-2em}
  \mu(\omega) = \sum_{N,N'} \mu_{NN'}(\omega),
\label{mu-sum}
\\&&\hspace*{-2em}
  \mu_{NN'}(\omega) = 
  n_N \int \mathcal{F}_N(v_z) \,\sigma_{NN'}(v_z,\omega) \dd{v_z} 
\nonumber\\&&
  - n_{N'} \int \mathcal{F}_{N'}(v_z') 
  \,\sigma_{NN'}(v_z,\omega) \dd{v_z'},
\label{muNN'}
\eea
where $\sigma_{NN'}(v_z,\omega)$ is the 
(free-free)
partial photoabsorption cross section
for ions in the Landau state $N$ having longitudinal velocity 
$v_z$ and going to the final state $N'$, and $\mathcal{F}_N(v_z)$
is the distribution of $v_z$ for such ions.\footnote{This 
is essentially the partial cross section given by 
Eq.~(\ref{sigmaNN'ff}), except that the latter assumes $v_z=0$.
To simplify notations, we suppress the subscript `$\alpha$' and 
the superscript `ff' in this section.}
The second term in \req{muNN'} represents stimulated emission
(treated as negative absorption), $v_z'$ is related
to $v_z$ by the energy conservation law
$
   \mpr v_z^2/2  + E_{N,\perp} +\hbar\omega
     = \mpr v_z'^2/2  + E_{N',\perp},
$
and the integration 
is performed over those $v_z$ for which this law can be satisfied.
In the case where $\mathcal{F}_N(v_z)=\mathcal{F}_{\mpr,T}(v_z)$
is the Maxwellian distribution (\ref{Maxwell})
with $T$ independent of $N$, \req{muNN'} can be 
written as
\beq
   \mu_{NN'}(\omega) = 
     n_N \sigma_{NN'}(\omega)
      \left[
      1-\frac{n_{N'}}{n_N}\,\mathrm{e}^{(N' - N) \betp - \hbar\omega/T}
       \right] ,
\label{muNN'1}
\eeq
where $\sigma_{NN'}(\omega) = 
\int \mathcal{F}_N(v_z) \,\sigma_{NN'}(v_z,\omega) \dd{v_z}$.
In LTE, Eqs.~(\ref{mu-sum})\,--\,(\ref{muNN'1}) yield 
\beq
\mu^{\rm LTE}(\omega) = \npr \sigma(\omega)\,(1-\mathrm{e}^{-\hbar\omega/T}),
\eeq
where $\npr\equiv \sum_N n_N$
and $\sigma(\omega)$ is the average photoabsorption cross section
of a proton:
$\sigma(\omega) = \sum_N f^p_N \sum_{N'} \sigma_{NN'}(\omega)$.

\begin{figure}%
\includegraphics[width=70mm]{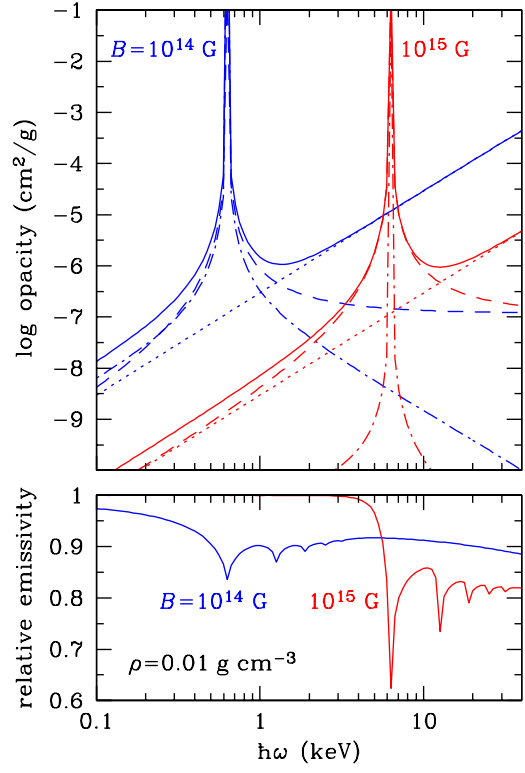}
\caption{Upper panel: The opacity components for the polarization
$\alpha=+1$, as functions of the photon energy,
for $\rho=0.01$ \gcc, $T=10^7$~K, and
$B=10^{14}$ and $10^{15}$~G (as marked
near the curves). Dashed lines -- ion scattering opacity,
dotted lines -- electron scattering opacity,
dot-dashed lines -- free-free absorption contribution,
solid lines -- the total. Lower panel: relative
emissivity, \req{relem}, for the same plasma parameters.
}
\label{fig:emi_a}
\end{figure}

\begin{figure}%
\includegraphics[width=70mm]{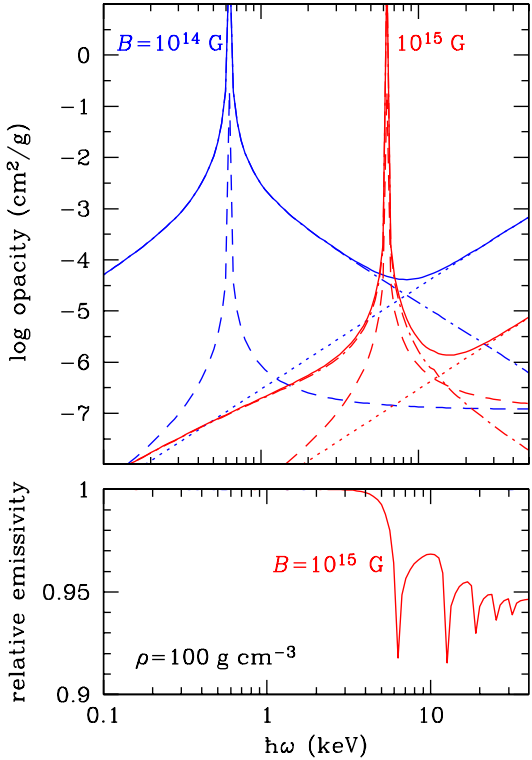}
\caption{The same as in Fig.~\ref{fig:emi_a}
but for $\rho=100$ \gcc.
In this case 
the relative emissivity for  the weaker field ($B=10^{14}$~G)
equals to one because of the LTE.
}
\label{fig:emi_b}
\end{figure}

The power of spontaneous emission of unit volume 
into $\dd\omega\,\dd\mathbf{n}$
is $j_\omega\,\dd\omega\,\dd\mathbf{n}$, where $j_\omega$ 
is the \emph{emission coefficient}.
Some authors (e.g., \citealt{Zhel}) call it emissivity
(whereas other authors, e.g., \citealt{RybickiLightman}
call emissivity the emission power
per unit mass). It
can be derived from the second Einstein relation (\ref{Ein-2})
and presented in the form
\bea&&\hspace*{-2em}
   j_\omega = \sum_{N'N} j_{\omega,N'N},
\label{j-sum}
\\&&\hspace*{-2em}
   j_{\omega,N'N} = \frac{\hbar\omega^3}{8\pi^3 c^2}
    n_{N'} \int \mathcal{F}_{N'}(v_z') 
  \,\sigma_{NN'}(v_z,\omega) \,\dd{v_z'} .
\label{j}
\eea
In the case of Maxwell--Boltzmann distribution of $v_z\low$ and $v_z'$,
using \req{muNN'1}
we obtain
\beq
   j_{\omega,N'N} = \frac{\hbar\omega^3}{8\pi^3 c^2}\,
          n_{N'}\, \sigma_{NN'}(\omega)
          \,\mathrm{e}^{(N' - N) \betp-\hbar\omega/T}
   \, .
\label{jNN'}
\eeq
In LTE, equations (\ref{mu-sum}), (\ref{j-sum}) and (\ref{jNN'})
reduce to the Kirchhoff law
(for each polarization mode)
\beq
j_\omega^{\rm LTE} = \mu^{\rm LTE}(\omega) B_\omega(T)/2.
\eeq
According to \req{jNN'}, the ratio of the emission coefficient 
to its LTE value (often also called emissivity) is 
\bea
   \frac{j_{\omega}}{j_{\omega}^{\mathrm{LTE}}}
   &=&
    \sum_{N'N} \frac{\mu_{NN'}(\omega)}
{\mu^{\rm LTE}(\omega)}
    \,\big(\mathrm{e}^{\hbar\omega/T} - 1\big)
\nonumber\\&&\times
   \bigg[\frac{n_N}{n_{N'}}\,\mathrm{e}^{(N - N') \betp + \hbar\omega/T} - 1
\bigg]^{-1} .
\label{relem}
\eea
We shall call ratio (\ref{relem}) \textit{relative emissivity}.

%%%%%%%%%%%%%%%%%%%%%%%%%%%%%%%%%%%%%%%%%%%%%%%%%%%%%%%%%%%%%%%%%%%%%
\subsection{Proton cyclotron line}

Figures \ref{fig:emi_a} and \ref{fig:emi_b}
show the opacities
     $\mu(\omega)/\rho$
 (upper panels) 
and relative
emissivities $j_{\omega}/j_{\omega}^{\mathrm{LTE}}$
(lower panels) 
as functions of the photon energy
for polarization $\alpha=+1$,
for two values of $\rho$ and two values of $B$.
At $\omega\ll\omcp$, 
the main contribution to the absorption and emission
of photons is given by the free-free processes
preserving $N$ ($N=N'$). 
In this case $j_\omega\approx j_\omega^\mathrm{LTE}$.
At higher $\omega$, transitions $N\to N'\neq N$
give a noticeable contribution to the photoabsorption.
In the absence of statistical equilibrium,
they result in a decrease of the relative emissivity
($j_\omega < j_\omega^\mathrm{LTE}$). 

The Coulomb logarithms for photoabsorption processes
$N\to N'$ strongly increase at $\omega\approx (N'-N)\omcp$.
At these frequencies the weight of such 
transitions increases, which causes weak spikes (pseudoresonances)
in the photoabsorption cross sections
(see \citealt{PC03}). 
   However, at the $\rho$, $T$, and $B$ values shown 
   in Figs.~\ref{fig:emi_a} and \ref{fig:emi_b},
   the resonant peaks of the Coulomb logarithms 
   $\Lambda_{0,N;\,0,N'}^{\mathrm{ff},+1}$ ($N'>N$)
   are not sufficiently high 
   to make them larger than $\Lambda_{0,N;\,0,N}^{\mathrm{ff},+1}$.
   Therefore the free-free absorption/emission processes with $N'=N$
   give the main contribution even at $\omega\approx (N'-N)\omcp$.
   Accordingly, these pseudoresonances are not very pronounced. They
are not visible
in the opacity curves in the upper panels
of Figs.~\ref{fig:emi_a} and \ref{fig:emi_b} because of
the logarithmic scale,
but the corresponding periodic decreases of the relative emissivity
at multiples of $\omcp$ are clearly seen in the lower panels.

%%%%%%%%%%%%%%%%%%%%%%%%%%%%%%%%%%%%%%%%%%%%%%%%%%%%%%%%%%%%%%%%%%%%%
\section{Generalization for other ions}
\label{sect:Z}

The formulae derived in the present paper for protons
can be generalized for other nuclei with arbitrary $A$ and $Z$.
If they have spin $\frac12$,
then in the formulae for the rates
of non-radiative collisions it is sufficient to replace
the mass $\mpr$ by $\mion=0.9928\,A\mpr$,
the number density $\npr$ by $\nion=n_e/Z$,
 the magnetic field parameter
 $\betp$ by $\beti \equiv \hbar\omci/ T = 73.9\,(Z/A)\,B_{14}/T_6$,
 and
to use for the Debye screening wave number 
the $Z$-dependent formula $\ks^2=4\pi(1+Z) n_e e^2/T$.
Furthermore, in \req{wpe} one should replace $I_{NN'}(t/2)$
by $I_{NN'}(t/2Z)$ and use \req{u'z1z2},
${u'}^2 = u^2+2Z(N-N')m_\ast/\mion$.
Also in \req{gam-pp} one should substitute $w(u_\pm)$
and $w^x(u_+,u_-)$ from Eqs.~(\ref{wN1N2}) (with $Z_1=Z_2=Z$)
and (\ref{wx}), and to use
${u'}^2 = u^2 + Z (N-N'+N_2-N_2')$ from \req{u'z1z2}.

Generalization for ions with different spin
is also straightforward, but more elaborate,
because it requires to rewrite \req{GppL}
with allowance for different projections of spin
on the magnetic field.

A possible generalization of the free-free cross section
 $\sigma^\mathrm{ff}$
   for $Z\neq1$ 
 is discussed at the end of Appendix~\ref{sect:ff}.

%%%%%%%%%%%%%%%%%%%%%%%%%%%%%%%%%%%%%%%%%%%%%%%%%%%%%%%%%%%%%%%%%%%%%
\section{Summary}

We have derived 
the general expressions 
for the rates of transitions between
     ion
 Landau 
levels
caused by non-radiative and radiative
    electron-ion
Coulomb collisions and non-radiative
     ion-ion
collisions.
We have also obtained (in Appendix~\ref{sect:ff})
the formulae for free-free photoabsorption cross sections
in strong magnetic fields with allowance for the electron
and
     ion (proton)
 quantization, which are much simpler
than the previously known ones.

On the base of the calculated transition rates we solved 
the equation of statistical equilibrium for protons
in a strong magnetic field. 
Considerable deviation from the Boltzmann distribution 
over the proton Landau levels occurs at densities 
$\rho\lesssim 0.1\,B_{14}^{3.5}$ \gcc.
At higher densities (lower magnetic fields)
non-LTE effects are negligible.
Conversely, at lower densities (higher $B$)
the excited proton states become depleted
because of radiative decay.

Nevertheless, even with strongly depleted
populations of the excited Landau states,
the emissivity of the fully ionized plasma
is not much suppressed relative to its LTE value.
This is because the main contribution in the
photoabsorption is given by transitions
which do not change Landau number $N$.
For such transitions, the relative population
of other levels is unimportant.

Although we have
     performed calculations only for
the proton-electron plasma,
generalization of our results to other ions
is rather straightforward (Sect.~\ref{sect:Z}).

All results in the present paper are
obtained in the 
Born
approximation.
A truncation used in older papers
to eliminate the divergence at small velocities, 
inherent to the Born approximation,
is now replaced by a smooth correction.
The thermally averaged
electron-proton non-radiative and radiative
transition rates are not sensitive to 
this correction. Proton-proton transition rates,
however, are sensitive, therefore
their more thorough examination 
beyond the Born approximation would be desirable.
Fortunately, at the considered physical conditions 
the proton-proton rates are not dominant,
thus we think that our main conclusions are sufficiently robust.

Our general expressions for the proton/ion Landau level transitions derived
in this paper will be useful for studying the possibility and the conditions 
of proton/ion cyclotron line formation in magnetar bursts. 
The radiation-dominated
bubble formed during the magnetar outbursts may be considered as a hot
($T\gtrsim 10$~keV), optically thick, rarefied medium embedded in a strong
magnetic field \citep{ThompsonDuncan05,WoodsThompson}. 
Within the bubble, vacuum polarization dominates the dielectric tensor 
and scattering dominates the opacity. The bubble may also contains
appreciable amount of ions ripped out of the NS surface during the outbursts. 
A neutron star atmosphere code, such as that developed in
\citet{HoLai01,HoLai03} and \citet{vanAdelLai06},
can be adapted to study radiative transfer in the bubble.
It may be that depending on the total bubble energy and the location of
energy release in the bubble, the characteristics of 
bubble radiation (such as ion cyclotron line strength,
line emission vs. absorption) are different. If so, the burst
spectra can provide a useful diagnostics for the energy dissipation
mechanisms of magnetar outbursts.

%%%%%%%%%%%%%%%%%%%%%%%%%%%%%%%%%%%%%%%%%%%%%%%%%%%%%%%%%%%%%%%%%%%%%

\section*{Acknowledgments}

A.P.\ thanks G\'erard Massacrier for enlightening discussions
of quantum effects on
spectral line formation, and theoretical astrophysics groups
of the Astronomy Department of Cornell University 
and of the Ecole Normale Sup\'erieure de Lyon for hospitality.
This work has been supported in part by NSF grant AST 0307252,
and {\it Chandra} grant TM6-7004X
(Smithsonian Astrophysical Observatory).
The work of A.P.\ is supported in part by FASI grant NSh-9879.2006.2
and RFBR grants 05-02-16245 and 05-02-22003.

%%%%%%%%%%%%%%%%%%%%%%%%%%%%%%%%%%%%%%%%%%%%%%%%%%%%%%%%%%%%%%%%%%%%%

%%%%%%%%%%%%%%%%%%%%%%%%%%%%%%%%%%%%%%%%%%%%%%%%%%%%%%%%%%%%%%%%%%%%%
\appendix
\section{Transition rates for Coulomb scattering}
\label{sect:w-general}

Here we present the non-relativistic formulae for the collision rates
of charged particles in an arbitrary magnetic field,
derived in the Born approximation.

%%%%%%%%%%%%%%%%%%%%%%%%%%%%%%%%%%%%%%%%%%%%%%%%%%%%%%%%%%%%%%%%%%%%%
\subsection{Scattering off a fixed Coulomb centre}
\label{sect:fixed}

%%%%%%%%%%%%%%%%%%%%%%%%%%%%%%%%%%%%%%%%%%%%%%%%%%%%%%%%%%%%%%%%%%%%%
\subsubsection{Derivation of the cross section for $Z=1$}

The motion of a charged particle in a magnetic field can be described 
by different sets of wave functions, corresponding to different choices
of the electromagnetic gauge \citep[e.g.,][]{LaLi-QM}. 
\citet{Ventura73} studied the 
scattering by a fixed Coulomb potential using the axially 
symmetric gauge of the vector potential $\bm{A}=\frac12\bm{B}\times\bm{r}$.
\citet{PavlYak} found that the Landau gauge
$(A_x,A_y,A_z) = (0, -By, 0)$ facilitates obtaining
a simpler representation of the scattering rate.
In the latter gauge, the `good' quantum numbers are the Landau number $N$
and the components $k_x$ and $k_z$ of the wave vector of the particle.
Assuming $Z=1$, one can write the coordinate wave function 
as
\beq
   \psi_{N,k_x,k_z}(\bm{r}) = (L_x L_z)^{-1/2} \mathrm{e}^{\mathrm{i} k_x x + \mathrm{i} k_z z}
   \hi_N(y+k_x\am^2),
\label{psi}
\eeq
where
\beq  
   \hi_N(y) = \frac{\exp(-y^2/2\am^2)}{\pi^{1/4}(2^N N! \, \am)^{1/2}}
   \,H_N(y/\am),
\label{chi}
\eeq
and 
$H_N(\xi)=(-1)^N {\rm e}^{\xi^2}{\rm d}^N{\rm e}^{-\xi^2}/{\rm d}\xi^N$
is a Hermite polynomial. The functions $\hi_N(y)$
are ortho-normalized and have the following nice property 
\citep{Klepikov,KaYak}:
\bea&&\hspace*{-2em}
   \int_{-\infty}^\infty \hi_{N}(y-q_x\am^2/2) \hi_{N'}(y+q_x\am^2/2) 
   \mathrm{e}^{\mathrm{i} q_y y} \dd y
\nonumber\\&&
   =  I_{N'N}(q_\perp^2 \am^2/2) \, \mathrm{e}^{\mathrm{i} (N'-N)\mathrm{arctan}(q_y/q_x)}.
\label{int-chi}
\eea
Here $q_\perp^2\equiv q_x^2+q_y^2$,
and $I_{N'N}$ is the Laguerre function \citep{SokTer},
defined as follows: if $N'-N\geq0$, then
\bea&&\hspace*{-2em}
   I_{N'N}(x) = \sqrt{\frac{N!}{N'!}} \,\mathrm{e}^{-x/2} x^{(N'-N)/2} L_N^{N'-N}(x)
\nonumber\\&&
   = \mathrm{e}^{-x/2} x^{(N'-N)/2}
       \sum_{l=0}^{N} \frac{\sqrt{N!N'!}\,(-x)^l}{l!\,(N-l)!\,(N'-N+l)!} ,
\label{Laguerre}
\eea
otherwise $I_{N'N}(x)=(-1)^{N-N'}I_{NN'}(x)$;
$L_N^s(x)$ is generalized Laguerre polynomial.

Relation (\ref{int-chi}) allows one to reduce
the matrix element of the transition 
$|Nk_x k_z\rangle \to |N'k_x'k_z'\rangle$
to the form
\bea&&\hspace*{-2em}
   M = \frac{1}{2\pi L_x L_z}\int_{-\infty}^\infty
   \mathrm{e}^{-iq_y(k_x+k_x')\am^2/2 + \mathrm{i} (N'-N)\mathrm{arctan}(q_y/q_x)}
\nonumber\\&&\times
     I_{N'N}(q_\perp^2\am^2/2) V_q \,\dd q_y ,
\label{M}
\eea
where $q_x = k_x'-k_x$, $q_z = k_z'-k_z$, and
\beq
   V_q = e^2 \int\dd\bm{r} \mathrm{e}^{\mathrm{i}\bm{q}\cdot\bm{r}}\,\frac{e^{-\ks r}}{r} = 
   \frac{4\pi e^2}{q^2+\ks^2}
\label{Vq}
\eeq
is the Fourier transform of the potential. In this paper we use the
screened Coulomb potential for the electron-ion
non-degenerate plasma,
for which $\ks$ equals the inverse Debye screening length.

Averaging the specific transition rate 
$(2\pi/\hbar)|M|^2 \delta(E'-E)$
(where $E$ and $E'$ is the initial and final energy)
over $k_x$ and summing it over $k_x'$ and $k_z'$
we obtain the transition rate per particle
on the level $N$ with initial longitudinal velocity $v_z=\hbar k_z/m$,
in a volume $V=L_x L_y L_z$:
\bea&&\hspace*{-2em}
   W_{NN'}^\mathrm{fix} 
   \equiv \frac{v_z \sigma_{NN'}^\mathrm{fix}(v_z)}{V} 
\nonumber\\&&
  =
  \frac{L_x L_z}{(2\pi)^2} \frac{\am^2}{L_y}
  \sum_\pm \int_{-\infty}^\infty \dd k_x \int_{-\infty}^\infty
   \dd k_x'
   \frac{2\pi}{\hbar} 
   |M|^2 \left(\frac{\dd E'}{\dd |k_z'|}\right)^{\!-1}
\nonumber\\&&
      = \frac{4\pi}{\tau_0} \frac{m}{\mel} \frac{\am^3}{V}
      \sum_\pm \frac{w_{NN'}^\mathrm{fix}(u_\pm)}{u'} ,
\label{WNN'fix}
\eea
where $\tau_0 = \hbar^3/e^4\mel$ is the atomic unit of time,
$\sigma_{NN'}^\mathrm{fix}(v_z)$ is an effective partial 
cross section,
\begin{subequations}
\label{wNN'fix}
\bea&&\hspace*{-2em}
   w_{NN'}^\mathrm{fix}(u_\pm) = \int_0^\infty 
   \frac{I_{N'N}^2(t/2)}{(t+u_\pm^2)^2 } \,\dd t ,
\\&&\hspace*{-2em}
      u_\pm = [(u\pm u')^2 + \us^2]^{1/2},
\quad
   u=|k_z|\am,
   \quad
   \us=\ks\am,
\\&&\hspace*{-2em}
   u' = |k_z'|\am = \sqrt{u^2 + 2 N - 2N'} ,
\eea
\end{subequations}
and $w^\mathrm{fix}_{NN'}$ should be set 
equal to zero when $u^2 < 2(N'-N)$.
It is easy to check that equations 
(\ref{WNN'fix}), (\ref{wNN'fix}) are equivalent
to equations (12), (13) of \citet{PavlYak}.

%%%%%%%%%%%%%%%%%%%%%%%%%%%%%%%%%%%%%%%%%%%%%%%%%%%%%%%%%%%%%%%%%%%%%
\subsubsection{Classical limit}

The function $w_{NN'}^\mathrm{fix}(u_\pm)$ belongs to a class
of integrals studied by \citet{P96}. 
According to his equation (B21), 
     based on \citet{KaYak},
in the semiclassical limit
\bea&&\hspace*{-2em}
   w_{NN'}^\mathrm{fix}(u_\pm) \approx 
   (1/2)\left(u_\pm^2/2 + N + N'\right)
\nonumber\\&&\times
    [(\sqrt{N'}-\sqrt{N})^2+u_\pm^2/2]^{-3/2}
 \label{wNN'appr}
\\&&\times
    [(\sqrt{N'}+\sqrt{N})^2+u_\pm^2/2]^{-3/2} 
\quad     (N\gg1,~N'\gg1).
\nonumber\eea
Using this approximation, one can demonstrate
 that equations (\ref{WNN'fix}),
(\ref{wNN'fix}) provide the correct classical cross section
in the limit $B\to 0$, where $N$ can be replaced 
by $(p_\perp\am/\hbar)^2/2$, $\bm{p}= m\bm{v}=\hbar\bm{k}$,
$p_\perp^2\equiv p_x^2+p_y^2$.
For example, consider a scattering event without screening ($\ks=0$), 
where the 
particle moves in the $xz$ plane 
at the angle $\theta$ to the $z$ axis before scattering and 
$\theta'=\theta+\alpha$
after scattering, $\theta<\pi/2$, and $\theta'<\pi/2$.
Taking into account that $p'=p$ 
and $ %\cos(\theta'-\theta)
(p_z'p_z\low-p_\perp' p_\perp\low)/p^2
=\cos\alpha = 1-2\sin^2\alpha/2$,
from \req{wNN'appr} we obtain
\beq
   w_{NN'}^\mathrm{fix}(u_-) = \left(\frac{\hbar}{\am}\right)^{\!4}
   \frac{1}{8p^3\,\sin^4\alpha/2}
   \frac{1-\cos\theta\cos\theta'}{1-\cos\alpha}.
\label{wNN'cl}
\eeq
When $N'\gg1$, a sum over $N'$
can be replaced by an integral. 
Then, according to equations (\ref{WNN'fix}) and (\ref{wNN'cl}),
the effective cross section for transitions
into a range of Landau levels
between $N'$ and $N'+\dd N'$ is
\bea&&\hspace*{-2em}
   \dd \sigma = \frac{4\pi}{\tau_0} \frac{m}{\mel} \frac{\am^3}{v_z}
   \frac{w(u_-)}{u'} \,\dd N'
   =
   \frac{4\pi}{\tau_0} \frac{m}{\mel} \frac{\am^4}{V\hbar}
   w(u_-) \frac{p_\perp'\dd p_\perp'}{p_z'}
\nonumber\\&&
   =
   \frac{\pi e^4 m \,(1-\cos\theta\cos\theta') \sin\theta'
      }{
   2Vp^3\,(1-\cos\alpha)\,\sin^4\alpha/2} 
   \,\dd\theta' .
\label{Wclass}
\eea
In the cylindrically symmetric case, 
when $\theta\to0$ and $\theta'\to\alpha$,
\req{Wclass} becomes
\beq
   \dd \sigma = \frac{e^4}{4 m^2 v_z^4}
   \,\frac{1}{\sin^4\alpha/2}\,\dd\Omega_\alpha,
\eeq
where $\dd\Omega_\alpha$ is a solid angle element.
This is the Rutherford formula.
 
%%%%%%%%%%%%%%%%%%%%%%%%%%%%%%%%%%%%%%%%%%%%%%%%%%%%%%%%%%%%%%%%%%%%%
\subsubsection{Cross section and average transition rate
for arbitrary $Z$}

For arbitrary charges of the scattered particle, $Ze$, 
and the Coulomb centre, $Z_0 e$, one should replace
 $\am \to \am|Z|^{-1/2}$ 
in equations (\ref{psi})\,--\,(\ref{M}) and $V_q \to Z_0 Z V_q$ in
     \req{M}.
Then 
\beq
      \sigma_{NN'}^\mathrm{fix}(v_z) = 
      \frac{4\pi\am^3}{v_z\tau_0}\, \frac{m}{\mel}
      Z_0^2 \sqrt{|Z|}
      \sum_\pm \frac{w_{NN'}^\mathrm{fix}(u_\pm)}{u'} ,
\label{WNN'fixZ}
\eeq
where $w_{NN'}^\mathrm{fix}$, $u_\pm$, and $u'$ are given 
by equations (\ref{wNN'fix}) with modified
scaling of $k_z$ and $k_z'$,
$u=|k_z|\am|Z|^{-1/2}$ and $u'=|k_z'|\am|Z|^{-1/2}$.
Besides, the $Z$-dependence of the Debye screening
parameter $\us=\ks\am$ should be taken into account.

If the velocities $v_z=\hbar k_z/\mion$ have Maxwellian distribution
(\ref{Maxwell}),
then from \req{WNN'fixZ} we obtain the probability
for one particle in the state $N$ in unit volume
to make a transition to the state $N'$
\beq
  \langle v_z \sigma^\mathrm{fix}_{NN'}(v_z) \rangle
   = \frac{4\sqrt{2\pi}}{\tau_0}
  \frac{m}{\mel}\,Z_0^2 \sqrt{|Z|}
       \, \am^3 \,
    \tilde\Lambda^\mathrm{fix}_{NN'} ,
\eeq
where $\tilde\Lambda^\mathrm{fix}_{NN'}
  \equiv \sqrt{\beta}\,\Lambda^\mathrm{fix}_{NN'}$,
\beq
  \Lambda^\mathrm{fix}_{NN'}
   = \int_0^\infty \frac{\dd u}{u'}
   \mathrm{e}^{- \beta u^2/2}\,\theta({u'}^2)\,
   \sum_\pm w^\mathrm{fix}_{NN'}(u_\pm) ,
\label{LNN'fix}
\eeq
$\beta = {\hbar|Z| eB}/{mcT}$.
The function $\theta({u'}^2)$ is the step function, equal to 1 
when $u^2 + 2(N-N')>0$ and 0 otherwise.

%%%%%%%%%%%%%%%%%%%%%%%%%%%%%%%%%%%%%%%%%%%%%%%%%%%%%%%%%%%%%%%%%%%%%
\subsubsection{Validity range and correction}
\label{sect:corr}

The integral in 
Eq.~(\ref{LNN'fix})
diverges when $N=N'$.
This behaviour, which is well known for
collision rates in the Born approximation in one dimension, 
means nothing but violation of this approximation for low
velocities of the colliding particles. 

A convenient parameter of the magnetic field strength at atomic scales
is
 $\gamma_B=b/(\alphaf Z Z_0)^2 
=\hbar^3 B/(m^2 c \,|Z|Z_0^2 e^3)$,
where $\alphaf$ is the fine-structure constant.
Born approximation is valid at $|k_z|,|k_z'|\gg|Z Z_0|e^2m/\hbar^2$,
that is at $u,u'\gg \gamma_B^{-1/2}
 $.
The most important effect of going
beyond Born approximation is the suppression of the
amplitude of the longitudinal part of the wave function
(the exponential in 
Eq.~[\ref{psi}]) near the Coulomb centre.
At $|k_z|\to0$ this amplitude becomes proportional
to $\sqrt{|k_z|}$. 
Quantitatively, when $\ln\gamma_B\gg1$,
the ratio of the square modulus of the amplitude 
 at $|k_z z| \,\gamma_B^{-1/4}\ll u \ll 1$
relative to its constant value at $u\gg1$
equals $C|k_z|\hbar^2/(m|ZZ_0|e^2)=Cu\sqrt{\gamma_B}$,
where $C=2\pi/\ln^2\gamma_B\,[1+O(1/\ln\gamma_B)]$
\citep{HaHo}.
Therefore, at $u\to0$ or $u'\to0$, $|M|^2$ in \req{WNN'fix}
becomes proportional to $u$ or $u'$, respectively,
which compensates the diverging factor $1/u'$ in \req{LNN'fix}.

In order to eliminate the divergence of collision integrals 
in Born approximation, similar to \req{LNN'fix},
previous authors \citep{PavlovPanov,KaYak} introduced
a cutoff at the lower limit of integration for $N=N'$.
Instead of the cutoff, we introduce weight function $g(u)g(u')$,
with
\beq
   g(u)=(1+\gamma_B^{-1} u^{-2})^{-1/2}.
\label{gBorn}
\eeq
Under the conditions $\gamma_B\gg1$ and $\beta\gamma_B\gg1$,
Born approximation is valid for most of the velocity values
that substantially contribute to the integral in \req{LNN'fix}.
The latter condition can be written as 
$T\gg Z_0^2 Z^2 e^4 m_\ast/\hbar^2$,
which is the usual condition of the applicability of Born approximation
in the non-magnetic case.

Apart from Born approximation,
which consists in neglecting the influence of the Coulomb
potential on the \emph{longitudinal} part of the wave function
(the exponential in Eq.~[\ref{psi}]),
we have also employed the adiabatic approximation,
which consists in neglecting perturbation of 
the \emph{transverse} part of the wave function
($\hi_N$ in Eq.~[\ref{psi}]).
For the continuum wave functions, both approximations are always valid
at $z\to\infty$, but may become inaccurate at the distances from the
Coulomb centre comparable to the Bohr radius. 
The adiabatic approximation remains sufficiently accurate
at small $z$ provided that the parameter $\gamma_B$,
introduced above, is large. Lowest-order perturbation corrections
and exact solution to the continuum wave functions 
in a strong magnetic field beyond the adiabatic 
approximation have been discussed, e.g., by \citet*{PPV}.

%%%%%%%%%%%%%%%%%%%%%%%%%%%%%%%%%%%%%%%%%%%%%%%%%%%%%%%%%%%%%%%%%%%%%
\subsection{Scattering of two charged particles}
\label{sect:2part}

%%%%%%%%%%%%%%%%%%%%%%%%%%%%%%%%%%%%%%%%%%%%%%%%%%%%%%%%%%%%%%%%%%%%%
\subsubsection{Scattering of different particles}
\label{sect:z1z2}

Let us consider Coulomb scattering of two particles
with charges $Z_i e$ ($i=1,2$).
In the Landau gauge (Sect.~\ref{sect:fixed}) their wave functions
are
\bea&&\hspace*{-2em}
   \psi_{N_i,k_{x,i},k_{z,i}}(\bm{r}_i) = (L_x L_z)^{-1/2}
    \mathrm{e}^{\mathrm{i} k_{x,i} x_i + \mathrm{i} k_{z,i} z_i}
\nonumber\\&&\times
   |Z_i|^{1/4}\,
   \hi_N\left(|Z_i|^{1/2} (y_i+k_{x,i}\am^2/Z_i)\right),
\label{psi^i}
\eea
where $\hi_N(y)$ is given by \req{chi}. The excitation energy 
of the two particles
is $E=\hbar eB c^{-1} (|Z_1|/m_1 + |Z_2|/m_2)
 + m_1 v_{z,1}^2/2 + m_2 v_{z,2}^2/2$,
where $\mion$ are masses of the particles,
and $v_{z,i}$ their longitudinal velocities.

Consider a transition in which the quantum numbers of the particles
change from $N_i,k_{x,i},k_{z,i}$ to $N_i',k_{x,i}',k_{z,i}'$ ($i=1,2$).
Wave functions (\ref{psi^i}) depend on $x$ and $z$ only through
the plane-wave exponential factor, which results in conservation of 
the $x$ and $z$ components of the total momentum 
in the matrix element of any potential which depends 
only on the relative position $\bm{r}_2-\bm{r}_1$
of the two particles: $k_{x1}+k_{x2}=k_{x1}'+k_{x2}'$,
$k_{z1}+k_{z2}=k_{z1}'+k_{z2}'$. 
Since $\hi_N$ does not depend on $k_{z,i}$,
we may choose the reference frame comoving with the centre of mass
in the $z$ direction, so that
$k_{z2}=-k_{z1}\equiv k_z$.
The number of final states in $\dd k_{x,1}'\, \dd k_{x,2}' \,\dd k_z'$ is
$(2\pi)^{-3} L_x^2 L_z \,\dd k_{x,1}' \,\dd k_{x,2}' \,\dd k_z'
  = (2\pi)^{-3} L_x^2 L_z 
  \,\dd k_{x,1}' \,\dd k_{x,2}' \,\dd E' \,m_\ast/\hbar^2|k_z'|$,
  where $m_\ast=m_1 m_2/(m_1+m_2)$ is the reduced mass.

Using Fourier decomposition 
of the interaction potential
$V(\bm{r}) = (2\pi)^{-3}\int\dd\bm{q}\,\mathrm{e}^{-\mathrm{i}\bm{q}\cdot\bm{r}}\,V_q$
(where $\bm{r} = \bm{r}_2-\bm{r}_1$),
and assuming $L_x$ and $L_z$ to be large, 
we can perform the integration over $x_1$, $q_x$, $z$, and $q_z$
in the matrix element $M$ for the transition 
$|N_1,N_2,k_{x,1},k_{x,2},k_z\rangle
 \to |N_1',N_2',k_{x,1}',k_{x,2}',k_z' \rangle$.
This integration fixes $q_x=k_{1,x}-k_{1,x}'$ and $q_z=k_z'-k_z$ in $V_q$.
Furthermore, using \req{int-chi},
we can perform integration over $y_1$ and $y_2$. Then we obtain
\bea&&\hspace*{-2em}
   M = \frac{1}{2\pi L_x^2 L_z\low}
   \int \dd x_2\,\mathrm{e}^{\mathrm{i}(q_x+k_{x,2}\low-k_{x,2}') x_2}
   \int\dd q_y \,V_q\,
\nonumber\\&&
   \times I_{N_1'N_1\low}\left(\frac{q_\perp^2\am^2}{2\,|Z_1|}\right)
    I_{N_2'N_2\low}\left(\frac{q_\perp^2\am^2}{2\,|Z_2|}\right)
\nonumber\\&&\times
   \exp\left[ \mathrm{i}\am^2 \left( \frac{k_{x,1}+k_{x,1}'}{2\,Z_1}
   - \frac{k_{x,2}+k_{x,2}'}{2\,Z_2} \right)\,q_y \right]
\nonumber\\&&\times
    \,\mathrm{e}^{ \mathrm{i}\,[(N_1'-N_1\low)\,\mathrm{sign}Z_1
    + (N_2'-N_2\low)\,\mathrm{sign}Z_2 ]\,\mathrm{arctan}(q_y/q_x) } .
\eea
By averaging $|M|^2$ over $k_{x,2}$ and summing
over $k_{x,2}'$, we arrive at
\bea&&\hspace*{-2em}
   \langle{|M|^2}\rangle_2 \equiv
    \int\frac{\am^2\,\dd k_{x,2}}{|Z_2|\,L_y} 
   \int\frac{L_x\,\dd k_{x,2}'}{2\pi}\,|M|^2
\nonumber\\&&\hspace*{-2em}
   = \frac{1}{2\pi L_x^2\,L_y\low\,L_z^2} \int\dd q_y\,|V_q|^2
   I_{N_1'N_1\low}^2\left(\frac{q_\perp^2\am^2}{2\,|Z_1|}\right)
    I_{N_2'N_2\low}^2\left(\frac{q_\perp^2\am^2}{2\,|Z_2|}\right) .
\nonumber\eea
This expression does not depend on $k_{x,1}$, and therefore
it does not require another averaging.
Finally, summation of the specific transition rate
$(2\pi/\hbar)\langle{|M|^2}\rangle_2\, \delta(E'-E)$
over $k_{x,1}'$ and $k_z'$ gives the partial transition
rate from $N_1,N_2$ to $N_1',N_2'$ for two 
particles in volume $V=L_x L_y L_z$
\bea&&\hspace*{-2em}
   W_{{N_1 N_2};\,{N_1' N_2'}} 
   \equiv V^{-1} v_z \,\sigma_{{N_1 N_2};\,{N_1' N_2'}}(v_z)
\nonumber\\&&
    = \frac{L_z m_\ast}{\hbar^3 |k_z'|}
   \,\sum_{\mathrm{sign}\,k_z'}
   \int\frac{L_x\,\dd k_x'}{2\pi}\,\langle|M|^2\rangle_2
\nonumber\\&&
    = \frac{4\pi}{\tau_0}\,\frac{m_\ast}{\mel}\,Z_1^2 Z_2^2\,\frac{\am^3}{V}
   \,\frac{1}{u'} \sum_\pm w_{N_1,N_2;\,N_1',N_2'}(u_\pm) ,
\label{WN1N2}
\eea
where
\bea&&\hspace*{-2em}
   w_{N_1,N_2;\,N_1',N_2'}(u)
    = \int_0^\infty \frac{\dd t}{(t+u^2)^2}
\nonumber\\&&\times
   \,I_{N_1'N_1\low}^2\left(\frac{t}{2\,|Z_1|}\right)
   I_{N_2'N_2\low}^2\left(\frac{t}{2\,|Z_2|}\right) .
\label{wN1N2}
\eea
Here $u_\pm = [(u\pm u')^2 + \us^2]^{1/2}$,
$u = |k_z|\am$, $\us=\ks\am$, and
\beq
   u' = \left[ u^2 + \frac{2\,m_2|Z_1|}{m_1+m_2}\,(N_1\low-N_1')
    + \frac{2\,m_1|Z_2|}{m_1+m_2}\,(N_2\low-N_2') \right]^{1/2} \!.
\label{u'z1z2}
\eeq
The last equation represents the energy conservation law.
The transition is energetically forbidden when the expression
in square brackets is negative. Note that we have redefined $u$ and $u'$
compared to the definition used in equations (\ref{WNN'fixZ})\,--\,(\ref{LNN'fix}): 
now $Z_i$ enter \req{u'z1z2}. Clearly, 
$w_{N_1,N_2;\,N_1',N_2'}(u)=w_{N_1',N_2;\,N_1,N_2'}(u)=w_{N_1,N_2';\,N_1',N_2}(u)$.
In addition, if $Z_1=Z_2$, then
$w_{N_1,N_2;\,N_1',N_2'}(u)=w_{N_2,N_1;\,N_2',N_1'}(u)$.

%%%%%%%%%%%%%%%%%%%%%%%%%%%%%%%%%%%%%%%%%%%%%%%%%%%%%%%%%%%%%%%%%%%%%
\subsubsection{Scattering of identical particles}
\label{sect:like}

The derivation of the transition rates for identical particles
can be patterned after Sect.~\ref{sect:z1z2}, but with initial
and final wave functions in the form
$
 \big[ \psi_{N_1,k_{x1},k_{z1}}(\bm{r}_1)\psi_{N_2,k_{x2},k_{z2}}(\bm{r}_2)
  \pm
\psi_{N_1,k_{x1},k_{z1}}(\bm{r}_2)\psi_{N_2,k_{x2},k_{z2}}(\bm{r_1})\big] / \sqrt{2} ,
$
where $\psi_{N,k_{x},k_{z}}(\bm{r})$ is given by \req{psi^i}.
The resulting partial transition rate from $N_1,N_2$
to $N_1',N_2'$, averaged over initial and integrated over final
$k_x$ values, is
\beq
   W_{N_1,N_2;\,N_1',N_2'}^\pm
    = 2\,(W_{N_1,N_2;\,N_1',N_2'} \pm W_{N_1,N_2;\,N_1',N_2'}^x),
\label{W-pm}
\eeq
where the sign $+$ ($-$) refers to the states with even (odd)
total spin, $W_{N_1,N_2;\,N_1',N_2'}$ is given by \req{WN1N2},
\beq
   W_{N_1,N_2;\,N_1',N_2'}^x = \frac{8\pi}{\tau_0}\,\frac{m_\ast}{\mel}
   \,\frac{|Z|^3}{u'} \, w_{N_1,N_2;\,N_1',N_2'}^x(u_-,u_+),
\eeq
and
\bea&&\hspace*{-2em}
   w_{N_1,N_2;\,N_1',N_2'}^x(u_-,u_+) =
    \int_0^\infty \frac{\dd t}{(t+u_+^2/|Z|)(t+u_-^2/|Z|)}
\nonumber\\&&\times
   I_{N_1'N_1\low}(t/2)\, I_{N_2'N_2\low}(t/2)\,
    I_{N_2'N_1\low}(t/2) \, I_{N_1'N_2\low}(t/2) .
\label{wx}
\eea
The latter function satisfies
symmetry relations 
$w_{N_1,N_2;\,N_1',N_2'}^x(u_-,u_+)
=w_{N_1,N_2;\,N_1',N_2'}^x(u_+,u_-)
=w_{N_2,N_1;\,N_1',N_2'}^x(u_-,u_+)
=w_{N_1,N_2;\,N_2',N_1'}^x(u_-,u_+)
=w_{N_1',N_2';\,N_1,N_2}^x(u_-,u_+)$.

%%%%%%%%%%%%%%%%%%%%%%%%%%%%%%%%%%%%%%%%%%%%%%%%%%%%%%%%%%%%%%%%%%%%%
\section{Cross sections of free-free photoabsorption}
\label{sect:ff}

In order to calculate a cross section of the
free-free absorption in a magnetic field,
it is important to take into account 
the the ion-electron centre of mass
 motion effects, even though $\mpr\gg\mel$.
 \citet{PC03} performed quantum calculations of this cross section 
 and demonstrated a correspondence to the classical
 dielectric tensor \citep{Ginzburg}. The result can be written as
\bea&&\hspace*{-2em}
   \sigma_\alpha^\mathrm{ff}(\omega)
   =
          \frac{4\pi e^2
          }{ 
     \mel c} \,
   \frac{\omega^2\,\nu_{\alpha}^{\mathrm{ff}}(\omega)
          }{
          (\omega+\alpha\omce)^2 (\omega-\alpha\omcp)^2
             +\omega^2 \tilde\nu_\alpha^2(\omega)},
\label{sigma-fit}
\\&&\hspace*{-2em}
\nu_{\alpha}^{\mathrm{ff}}(\omega) =
    \sum_{n,N} f^e_n f^p_N \sum_{n',N'}
\nu_{n,N;\,n',N'}^{\mathrm{ff},\alpha}(\omega) ,
\label{nu-ff}
\\&&\hspace*{-2em}
     \tilde\nu_\alpha(\omega) =
     \left(1+\alpha\,\frac{\omce}{\omega}\right)
     \nu_p(\omega)
     + \left(1-\alpha\,\frac{\omcp}{\omega}\right)
     \nu_e(\omega) + \nu_{\alpha}^{\mathrm{ff}}(\omega).
\nonumber\\
\label{damp}\eea
Here $\alpha=0,\pm1$ is the polarization index,
$\nu_{\alpha}^{\mathrm{ff}}$ is the effective
frequency of the electron-ion collisions
for a given photon frequency $\omega$,
$f^p_N$  and $f^e_n$
are the fractions of the protons and electrons in Landau states
$N$ and $n$, 
$\sigma_{\alpha;\,n,N;\,n',N'}^\mathrm{ff}(\omega)$
is a partial photoabsorption cross section for
a transition in which the electron and proton
change their Landau quantum numbers from $n$ to $n'$
and from $N$ to $N'$, respectively.
Finally, $\nu_p$ and $\nu_e$ in \req{damp} are 
effective damping factors for protons and electrons,
respectively, not related to the electron-proton collisions
(for example, \citet{Ginzburg} considers collisions
of electrons and protons with molecules).
The derivation of the damping factor (\ref{damp})
from the complex dielectric tensor of the plasma
assumes $\nu_{e}\ll\omce$, $\tilde\nu_\alpha\ll\omce$,
and $\nu_p\ll\omcp$.

\citet{PC03} calculated $\nu_{n,N;\,n',N'}^{\mathrm{ff},\alpha}(\omega)$
assuming LTE.
Following their approach without LTE, however retaining the Maxwell
longitudinal distributions (\ref{Maxwell}),
we present the result as follows:
\beq
   \nu_{n,N,n',N'}^{\mathrm{ff},\alpha}(\omega) = 
        \frac{4}{3}\,\sqrt{\frac{2\pi}{\mel T}}\,
          \frac{n_e\, e^4}{\hbar \omega}\,
 \Lambda_{n,N;\,n',N'}^\mathrm{ff,\alpha}(\beta_\ast,\omega/\omega_\ast) ,
\label{Gamma-nNn'N'}
\eeq
where for $\alpha=0$
\bea&&\hspace*{-2em}
   \Lambda_{n,N;\,n',N'}^\mathrm{ff,0}(\beta_\ast,\omega/\omega_\ast)
   = \frac32 \int_0^\infty \frac{\dd u}{u'}\,\mathrm{e}^{-\beta_\ast u^2/2}
   \,\theta({u'}^2)\,
\nonumber\\&&\times
    \sum_\pm (u'\pm u)^2\, w_{n,N;\,n',N'}^{(0)}(u_\pm),
\label{Lambda0}
\eea
and for $\alpha=\pm1$
\bea&&\hspace*{-2em}
   \Lambda_{n,N;\,n',N'}^\mathrm{ff,\alpha}(\beta_\ast,\omega/\omega_\ast)
   = \frac32 \int_0^\infty \frac{\dd u}{u'}\,\mathrm{e}^{-\beta_\ast u^2/2}
   \,\theta({u'}^2)\,
\nonumber\\&&\times
    \sum_\pm \bigg[ \frac{m_\ast^2}{\mel^2}
    \left( 1-\alpha\frac{\omcp}{\omega} \right)^{\!2}
       w_{n,N;\,n',N'}^{e,\alpha}(u_\pm)
\nonumber\\&&\!\!
   + \frac{2\mpr\mel}{(\mpr+\mel)^2}
    \left( 1-\alpha\frac{\omcp}{\omega} \right)
    \left( 1+ \alpha\frac{\omce}{\omega} \right)
       w_{n,N;\,n',N'}^{x,\alpha}(u_\pm)
\nonumber\\&&
   + \frac{m_\ast^2}{\mpr^2}
    \left( 1+\alpha\frac{\omce}{\omega} \right)^{\!2}
       w_{n,N;\,n',N'}^{\mathrm{i},\alpha}(u_\pm)
       \bigg].
\label{Lambda-gen}
\eea
Here 
$\beta_\ast = \hbar\omega_\ast/T = \hbar eB/m_\ast c T
  = \betp\mpr/m_\ast$,
and the arguments $u_\pm = [(u\pm u')^2 + \us^2]^{1/2}$,
$u = |k_z|\am$, $\us=\ks\am$, and
\beq
   u' = \left[ u^2 + \frac{2m_\ast}{\mpr}\,(N-N')
    + \frac{2m_\ast}{\mel}\,(n-n')
    + \frac{2\omega}{\omega_\ast}
     \right]^{1/2}
\label{u'ff}
\eeq
have the same meaning 
as in Appendix~\ref{sect:w-general}.
The functions $w(u_\pm)$ are defined as
\begin{subequations}
\label{w-detailed}
\bea&&\hspace*{-2em}
   w_{n,N;\,n',N'}^{e,+1}(u) = \int_0^\infty \rho\,\dd\rho
     \Big[ \sqrt{n'+1}\,\tilde{v}_{n,s,n'+1,s'-1}(\rho,u\sqrt{2})
\nonumber\\&&
     -\sqrt{n}\,\tilde{v}_{n-1,s+1,n',s'}(\rho,u\sqrt{2}) \Big]^2,
\\&&\hspace*{-2em}
   w_{n,N;\,n',N'}^{p,+1}(u) = \int_0^\infty \rho\,\dd\rho
     \Big[ \sqrt{N'}\,\tilde{v}_{n,s,n',s'-1}(\rho,u\sqrt{2})
\nonumber\\&&
     -\sqrt{N+1}\,\tilde{v}_{n,s+1,n',s'}(\rho,u\sqrt{2}) \Big]^2,
\\&&\hspace*{-2em}
   w_{n,N;\,n',N'}^{x,+1}(u) = \int_0^\infty \rho\,\dd\rho
     \Big[ \sqrt{n'+1}\,\tilde{v}_{n,s,n'+1,s'-1}(\rho,u\sqrt{2})
\nonumber\\&&
     -\sqrt{n}\,\tilde{v}_{n,s+1,n',s'}(\rho,u\sqrt{2}) \Big]
\,
     \Big[ \sqrt{N'}\,\tilde{v}_{n,s,n',s'-1}(\rho,u\sqrt{2})
\nonumber\\&&
     -\sqrt{N+1}\,\tilde{v}_{n,s+1,n',s'}(\rho,u\sqrt{2}) \Big],
\eea
\end{subequations}\noindent
where $s = N-n$ and $s' = N'-n'$ are the relative proton-electron
orbital quantum numbers, and
$\tilde{v}_{n,s,n',s'}(\rho,x)$ are the scaled Fourier transforms
of the effective potentials defined in Appendix~B
of \citet{PC03}.\footnote{In equation (B7)
of \citet{PC03} the factor
$\sqrt{\tilde{s}\tilde{s}'}$ (a typo)
must be $\sqrt{\tilde{s}!\tilde{s}'!}$.}
Due to the symmetry properties of these potentials
we have $w_{n,N;\,n',N'}^{(e,p,x),-1}(u) = w_{N,n;\,N',n'}^{(e,p,x),+1}(u)$.
Finally,
\beq
   w_{n,N;\,n',N'}^{(0)}(u) = \int_0^\infty \rho\,\dd\rho
     \,\tilde{v}_{n,s,n',s'}^2(\rho,u\sqrt{2}) .
\label{w0}
\eeq

One can demonstrate that the integrals (\ref{w-detailed})
are symmetric with respect to interchange of their 
 indices $n,n'$ or $N,N'$, and moreover, all of them coincide
with one another.
It follows that \req{Lambda-gen} simplifies to
\bea&&\hspace*{-2em}
   \Lambda_{n,N;\,n',N'}^\mathrm{ff,\pm1}(\beta_\ast,\omega/\omega_\ast)
   = \frac32 \int_0^\infty \frac{\dd u}{u'}\,\mathrm{e}^{-\beta_\ast u^2/2}
   \,\theta({u'}^2)\,
\nonumber\\&&\times
    \left[ w_{n,N;\,n',N'}^{(1)}(u_+) + w_{n,N;\,n',N'}^{(1)}(u_-) \right] ,
\label{Lambda1}
\eea
where $w_{n,N;\,n',N'}^{(1)}(u)$ is any 
of the integrals (\ref{w-detailed}).

The integral (\ref{Lambda1}) diverges at
$\omega\to \omcp(N'-N)+\omce(n'-n)$, which
is caused by the failure of Born approximation
at slow electron-proton relative motion. 
As discussed in Sect.~\ref{sect:corr},
this failure can be cured by introducing 
correction factors (\ref{gBorn}) under the integral.

The functions 
$w_{n,N;\,n',N'}^{(0,1)}(u)$ can be presented as
\beq
       w_{nN;\,n'N'}^{(\alpha)}(u)
    = \frac12 \int_0^\infty \frac{t^{|\alpha|}\,\dd t}{(t+u^2/2)^2}
   I_{n',n}^2(t)\,
   I_{N'N}^2(t) ,
\label{wN1N2a}
\eeq
This alternative representation can be obtained
by passing from the cylindrical to Landau gauge
in the derivation of the free-free cross section 
(Appendix B of \citealt{PC03}),
taking into account \req{int-chi} and recurrence
relation $L_n^s(x)-L_{n-1}^s(x)=L_n^{s-1}(x)$.

We see that
$w_{n,N;\,n',N'}^{(0)}(u)$ 
coincides with $w_{n,N;\,n',N'}(u)$
given by \req{wN1N2} at $Z_1=Z_2=1$,
Thus, for a given initial ($n,N$) and final ($n',N'$) Landau
quantum numbers,
the effective rates of the electron-proton
radiative and non-radiative transitions
are determined, for any $B$, $T$, and $\omega$,
by one-dimensional integrals involving
just two universal functions
$w_{n,N;\,n',N'}^{(0,1)}(u)$.
These functions are smooth and monotonically decreasing.
At $u\ll1$ they tend to constants,
except for the following cases:
(i) if $n'=n$ and $N'=N$, then
$w_{n,N;\,n',N'}^{(1)}(u)\sim \ln u$
and $w_{n,N;\,n',N'}^{(0)}(u)\sim u^{-2}$;
(ii) if $n'=n$ and $N'=N\pm1$ (or $N'=N$ and $n'=n\pm1$),
then $w_{n,N;\,n',N'}^{(0)}(u)\sim n_\ast\ln u$,
where $n_\ast=\max(N,N')$ (or $n_\ast=\max(n,n')$, respectively).
At $u\gg1$, $w_{n,N;\,n',N'}^{(\alpha)}(u)\propto u^{-4}$.

The approximation of infinitely
massive ions \citep{PavlovPanov}
is reproduced by setting $\omcp=0$ in \req{sigma-fit}
and replacing 
$I_{N'N}^2(t)$ by $\sum_{N'=0}^\infty I_{N'N}^2(t) = 1$ 
in \req{wN1N2a}. 

 Since \req{sigma-fit} is classical 
(although the factors $\nu^\mathrm{ff}_\alpha$
 and $\tilde\nu_\alpha$ need quantum calculation),
we may extend it to $Z>1$. In this case the right-hand side
of \req{Gamma-nNn'N'} 
should be multiplied by $Z^2$, and the 
coefficient
$\Lambda_{n,N;\,n',N'}$ will be different.
By analogy with the general case of non-radiative Coulomb
collisions, considered in Sect.~\ref{sect:z1z2},
the latter difference consists in replacing of
$I_{NN'}^2(t)$ by $I_{NN'}^2(t/Z)$ in \req{wN1N2a}
and multiplying $(N-N')$ by $Z$ in \req{u'ff}.
This simple generalization to the $Z\neq1$ case is possible only
in the adiabatic and Born approximations,
which we use in this paper. 
Beyond these approximations,
the effects of centre-of-mass motion of an ion-electron
system affect the initial and final wave functions
in a non-trivial way. Continuum wave functions with allowance for
these effects have been so far calculated only
for $Z=1$ \citep{PP97}.


\begin{thebibliography}{99}

\bibitem[Araya \& Harding(1999)]{ArayaHarding}
Araya R.~A., Harding A.~K., 1999, ApJ, 517, 334

\bibitem[Araya-G\'ochez \& Harding(2000)]{ArayaG-Harding}
Araya-G\'ochez R.~A., Harding A.~K., 2000, ApJ, 544, 1067
% Cyclotron-Line Features from Near-critical Fields. II. On the Effect
% of Anisotropic Radiation Fields

\bibitem[Baring et al.(2005)Baring, Gonthier \& Harding]{Baring-ea05}
Baring M.~G., Gonthier P.~L.,  Harding A.~K., 2005,
% ``Spin-dependent cyclotron decay rates in strong magnetic fields,''
ApJ, 630, 430%--440

\bibitem[Burnard et al.(1988)Burnard, Klein \& Arons]{BKA88} 
Burnard D.~J., Klein R.~I., Arons, J., 1988, ApJ, 324, 1001

\bibitem[Canuto et al.(1971)Canuto, Lodenquai \& Ruderman]{Canuto-ea}
Canuto V., Lodenquai J.,  Ruderman M., 1971,
Phys.\ Rev. D, 3, 2303

\bibitem[Cohen-Tannoudji et al.(1998)Cohen-Tannoudji, Dupont-Roc \& Grynberg]{Cohen-T-ea}
Cohen-Tannoudji C., Dupont-Roc J.,  Grynberg G., 1998,
Atom-Photon Interactions:
Basic Processes and Applications.
Wiley-VCH, Berlin

\bibitem[Daugherty \& Ventura(1977)]{DV77}
Daugherty J.~K.,  Ventura J., 1977,
% ``Cyclotron lines in the Her X-1 spectrum: Structure and higher harmonics,''
A\&A, 61, 723%--727

\bibitem[Daugherty \& Ventura(1978)]{DV78}
Daugherty J.~K.,  Ventura J., 1978,
% ``Absorption of radiation by electrons in intense magnetic fields,''
Phys.\ Rev. D, 18, 1053%--1067

\bibitem[Gavriil et al.(2002)Gavriil, Kaspi \& Woods]{GKW02} 
Gavriil F.~P., Kaspi V.~M., Woods P.~M., 2002, Nat, 419, 142
% (astro-ph/0209202)
% Magnetar-like X-ray bursts from an AXP

\bibitem[Ginzburg(1970)]{Ginzburg}
Ginzburg V.~L., 1970,
The Propagation of Electromagnetic Waves in Plasmas,
2d ed. Pergamon, London

\bibitem[Gnedin \& Sunyaev(1974)]{GnedinSunyaev74}
Gnedin Yu.~N.,  Sunyaev R.~A., 1974,
A\&A, 36, 379

\bibitem[Haberl et al.(2004)]{Haberl-ea04} 
Haberl F.\ et al., % , Motch C., Zavlin V.~E., Reinsch K., Gaensicke B.~T., 
% Cropper M., Schwope A.~D., Turolla R., Zane S., 
2004, A\&A, 424, 635 
% The isolated neutron star X-ray pulsars RX J0420.0-5022 and 
% RX J0806.4-4123: new X-ray and optical observations

\bibitem[Hasegawa \& Howard(1961)]{HaHo}
Hasegawa H.,  Howard R.~E., 1961,
J.\ Phys.\ Chem.\ Solids, 21, 179

\bibitem[Heindl et al.(2004)]{Heindl-ea04}
Heindl W.~A., Rothschild R.~E., Coburn W., Staubert R., 
Wilms J., Kreykenbohm I., Kretschmar P., 2004, 
% ``Timing and Spectroscopy of Accreting X-ray Pulsars: the State of Cyclotron Line Studies''
in 
Kaaret P., Lamb F.~K., Swank J.~H., eds,
X-ray Timing 2003: Rossi and Beyond,
AIP Conference Proceedings, Vol.~714. AIP, Melville, NY, p.~323%--330
% (astro-ph/0403197)

\bibitem[Ho \& Lai(2001)]{HoLai01} 
Ho W.~C.~G., Lai D., 2001, 
%``Atmospheres and Spectra of Strongly Magnetized Neutron Stars'', 
MNRAS, 327, 1081

\bibitem[Ho \& Lai(2003)]{HoLai03} 
Ho W.~C.~G., Lai D., 2003, MNRAS, 338, 233

\bibitem[Ho \& Lai(2004)]{HoLai04} 
Ho W.~C.~G., Lai D., 2004, ApJ, 607, 420

\bibitem[Ibrahim et al.(2003)Ibrahim, Swank \& Parke]{ISP03} 
Ibrahim A.~L., Swank J.~H., Parke W., 2003, ApJ, 584, L17
%(astro-ph/0210515)
%New evidence for proton cyclotron line from SGR 1806-20

\bibitem[Jackson(1975)]{Jackson}
Jackson J.~D., 1975,
Classical Electrodynamics,
2nd ed. Wiley, New York

\bibitem[Juett et al.(2002)]{Juett-ea02}
Juett A.~M., Marshall H.~L., Chakrabarty D., Schulz N.~S., 2002, ApJ, 
568, L31

\bibitem[Kaminker \& Yakovlev(1981)]{KaYak}
 Kaminker A.~D., Yakovlev D.~G., 1981,
 % ``Description of a relativistic electron
 % in a quantizing magnetic field.
 % Transverse transport coefficients of an electron gas,''
 Theor. Math. Phys., 49, 1012%--1020.

\bibitem[Klepikov(1954)]{Klepikov}
Klepikov N.~P., 1954,
Zh.\ Eksp.\ Teor.\ Fiz., 26, 19 (in Russian)

\bibitem[Kulkarni et al.(2003)]{Kulkarni-ea03} 
Kulkarni S.~R., Kaplan D.~L., Marshall H.~L., Frail D.~A., 
Murakami T., Yonetoku D., 2003, ApJ, 585, 948
% The quiescent counterpart of SGR 0526-66

% \bibitem[Lai \& Ho(2002)]{LaiHo02} 
% Lai D., Ho W.~C.~G., 2002, ApJ, 566, 373

\bibitem[Lai \& Ho(2003)]{LaiHo03} 
Lai D., Ho W.~C.~G., 2003, ApJ, 588, 962

\bibitem[Lamb \& Masters(1979)]{LambMasters}
Lamb D.~Q.,  Masters A.~R., 1979,
ApJ, 234, L117.

\bibitem[Lamb et al.(1990)Lamb, Wang \& Wasserman]{LWW90} 
Lamb D.~Q., Wang J.~C.~L., Wasserman I., 1990, ApJ, 363, 670

\bibitem[Landau \& Lifshitz(1976)]{LaLi-QM}
Landau L.~D.,  Lifshitz E.~M., 1976,
{Quantum Mechanics}.
Pergamon, Oxford

\bibitem[Langer(1981)]{Langer}
Langer S.~H., 1981,
% ``Collisional excitation of electron Landau levels in strong magnetic fields,''
Phys.\ Rev.\ D, 23, 328;
%--346
erratum: Phys.\ Rev.\ D, 25, 1157 (1982)

\bibitem[McLaughlin et al.(2003)]{McLaughlin-ea03} 
McLaughlin M.~A. %, Stairs I.~H., Kaspi V.~M., 
et al., 2003, ApJ, 591, L135

\bibitem[Melrose \& Zheleznyakov(1981)]{MZh}
Melrose D.~B.,  Zheleznyakov V.~V., 1981,
% ``Quantum theory of cyclotron emission and the X-ray line in Her X-1,''
A\&A, 95, 86%--93

\bibitem[M\'esz\'aros(1992)]{Mesz}
M\'esz\'aros P., 1992,
High-Energy Radiation from Magnetized Neutron Stars.
Univ.~of Chicago Press, Chicago

\bibitem[M\'{e}sz\'{a}ros \& Nagel(1985)]{MN85}
M\'{e}sz\'{a}ros P., Nagel, W., 1985, ApJ, 299, 138

\bibitem[Miller et al.(1987)Miller, Salpeter \& Wasserman]{MSW87}
Miller G.~S., Salpeter E.~E.,  Wasserman I., 1987,
% ``Deceleration of infalling plasma in the atmospheres of accreting neutron stars. I. Isothermal atmospheres,''
ApJ, 314, 215%--233

\bibitem[Nagel \& Ventura(1983)]{NagelVentura}
Nagel W.,  Ventura J., 1983,
A\&A, 118, 66

\bibitem[Patel et al.(2003)]{Patel-ea03} 
Patel S.~K., Kouveliotou C., Woods P., et al., 2003, ApJ, 587, 367
% Chandra observations of AXP 4U0142+61

\bibitem[Pavlov \& Panov(1976)]{PavlovPanov}
Pavlov G.~G.,  Panov A.~N., 1976,
% Zh.\ Eksper.\ Teor.\ Fiz. 71, 572
Sov.\ Phys.--JETP, 44, 300

\bibitem[Pavlov \& Yakovlev(1976)]{PavlYak}
Pavlov G.~G., Yakovlev D.~G., 1976,
% ``Coulomb deceleration of fast protons in a strong magnetic field,''
% Zh.\ Eksper.\ Teor.\ Fiz. 70, 753
%--767
Sov.\ Phys.--JETP, 43, 389%--396

\bibitem[Pavlov et al.(1991)]{Pavlov-ea91}
Pavlov G.~G., Bezchastnov V.~G., M\'esz\'aros P.,  Alexander S.~G., 1991,
% ``Radiative widths and splitting of cyclotron lines in superstrong magnetic fields,''
ApJ, 380, 541%--549

\bibitem[Pavlov et al.(1995)]{Pavlov95}
Pavlov G.~G., Shibanov Yu.~A., Zavlin V.~E.,  Meyer R.~D., 1995,
%% ``Neutron Star Atmospheres'',
in Alpar M.~A., Kizilo\u{g}lu \"U.,  van Paradijs J., eds,
The Lives of the Neutron Stars,
 NATO ASI Ser.~C, vol.~450.
Kluwer, Dordrecht, p.~71

\bibitem[Potekhin(1996)]{P96}
Potekhin A.~Y., 1996,
A\&A, 306, 999

\bibitem[Potekhin \& Chabrier(2003)]{PC03}
Potekhin A.~Y.,  Chabrier G., 2003,
ApJ, 585, 955

\bibitem[Potekhin \& Pavlov(1997)]{PP97}
Potekhin A.~Y.,  Pavlov G.~G., 1997,
ApJ, 483, 414

\bibitem[Potekhin et al.(1997)Potekhin, Pavlov \& Ventura]{PPV}
Potekhin A.~Y., Pavlov G.~G.,  Ventura J., 1997,
A\&A 317, 618

\bibitem[Potekhin et al.(2004)]{PLCH04} 
Potekhin A.~Y., Lai D., Chabrier G., Ho W.~C.~G., 2004, 
ApJ, 612, 1034

\bibitem[Potekhin et al.(2005)]{cospar04}
Potekhin A.~Y., Lai D., Chabrier G., Ho W.~C.~G., 2005,
Adv.\ Sp.\ Res., 35, 1158

\bibitem[Rea et al.(2003)]{Rea-ea03}
Rea N., Israel G.~L., Stella L., Oosterbroek T., Mereghetti S.,
Angelini L., Campana S., Covino S., 2003,
ApJ, 586, L65

\bibitem[Rea et al.(2005)]{Rea-ea05}
Rea N.,  Oosterbroek T., Zane S., Turolla R., M\'endez M.,
Israel G.~L., Stella L., Haberl F., 2005, MNRAS, 361, 710

\bibitem[Rybicki \& Lightman(1979)]{RybickiLightman}
Rybicki G.~B.,  Lightman A.~P., 1979,
Radiative Processes in Astrophysics.
Wiley, New York

\bibitem[Sokolov \& Ternov(1986)]{SokTer}
Sokolov A.~A.,  Ternov I.~M., 1986,
Radiation from Relativistic Electrons,
2d ed. AIP, New York

\bibitem[Storey \& Melrose(1987)]{StoreyMelrose}
Storey M.~C.,  Melrose D.~B., 1987,
% ``Collisions in strong magnetic fields,''
Aust.\ J.\ Phys., 40, 89%--107

\bibitem[Strohmayer \& Ibrahim(2000)]{StrohmayerIbrahim} 
Strohmayer T.~E., Ibrahim A.~I., 2000, ApJ, 537, L111

\bibitem[Terada et al.(2006)]{Terada-ea06}
Terada, Y., et al. 2006, ApJ, 648, L139

\bibitem[Thompson \& Duncan(1995)]{ThompsonDuncan05}
Thompson C., Duncan R.~C., 1995, MNRAS, 275, 255

\bibitem[Thompson \& Duncan(1996)]{ThompsonDuncan06}
Thompson C., Duncan R.~C., 1996, ApJ, 473, 322

\bibitem[Tiengo et al.(2005)]{Tiengo-ea}
Tiengo A., Mereghetti S., Turolla R., Zane S., Rea N., Stella L., Israel G. L.,
2005, 
A\&A, 437, 997
%Three XMM-Newton observations of the anomalous X-ray pulsar 1E
%1048.1-5937: Long term variations in spectrum and pulsed fraction

\bibitem[Tr\"umper et al.(1978)]{Trumper-ea78}
Tr\"umper J., Pietsch W., Reppin C., 
Voges W., Staubert R., Kendziorra E., 1978,
%``Evidence for strong cyclotron
%line emission in the hard X-ray spectrum of Hercules X-1,''
ApJ, 219, L105%--L110.

\bibitem[van Adelsberg \& Lai(2006)]{vanAdelLai06} 
van Adelsberg M., Lai D., 2006, MNRAS, in press (astro-ph/0607168)

\bibitem[van Kerkwijk et al.(2004)]{Kerkwijk-ea04} 
van Kerkwijk M.~H., Kaplan D.~L., Durant M., Kulkarni S.~R., Paerels F.,
2004, ApJ, 608, 432

\bibitem[van Kerkwijk \& Kaplan(2006)]{Kerkwijk06} 
van Kerkwijk M.~H., Kaplan D.~L., 2006, ApSS, in press (astro-ph/0607320)

\bibitem[Ventura(1973)]{Ventura73}
Ventura J., 1973,
% ``Collision frequency and Coulomb scattering in an intense magnetic field,''
Phys.\ Rev. A, 8, 3021%--3031

\bibitem[Ventura(1979)]{Ventura79}
Ventura J., 1979,
Phys.\ Rev. D, 19, 1684

\bibitem[Ventura et al.(1979)Ventura, Nagel \& M\'esz\'aros]{Ventura-ea79}
Ventura J., Nagel W.,  M\'esz\'aros P., 1979,
ApJ, 233, L125

\bibitem[Wang et al.(1993)Wang, Wasserman \& Lamb]{WWL93}
Wang J.~C.~L., Wasserman I., Lamb D.~Q., 1993, ApJ, 414, 815

\bibitem[Wasserman \& Salpeter(1980)]{WassermanSalpeter}
Wasserman I., Salpeter E.~E., 1980, ApJ, 241, 1107

\bibitem[Woods \& Thompson(2005)]{WoodsThompson} 
Woods P.~M., Thompson C., 2005, 
in Lewin W.H.G., van der Klis M., eds,
Compact Stellar X-ray Sources,
Cambridge Univ.\ Press, Cambridge

\bibitem[Zane et al.(2001)]{Zane-ea01}
Zane S., Turolla R., Stella L., Treves A., 2001,
ApJ, 560, 384

\bibitem[Zheleznyakov(1996)]{Zhel}
Zheleznyakov V.~V., 1996,
Radiation in Astrophysical Plasmas.
Kluwer, Dordrecht

\end{thebibliography}
\end{document}